\documentclass{jfm}
\usepackage{graphicx}
\usepackage{epstopdf, epsfig}
\usepackage[utf8]{inputenc}
\usepackage[english]{babel}
\usepackage[T1]{fontenc}
\usepackage{lmodern}
\usepackage[dvipsnames]{xcolor}
\usepackage[colorlinks=true,linkcolor=blue,citecolor=blue,filecolor=blue,urlcolor=blue]{hyperref}
\usepackage[normalem]{ulem}

\usepackage{amssymb}
\usepackage{bm}
\usepackage{lastpage} 
\usepackage{setspace}
\usepackage{fancyhdr} 
\usepackage{graphicx}
\usepackage{tikz}
\usetikzlibrary{patterns}

\usepackage{siunitx}
\usepackage{amsmath}
\usepackage{chemformula} 
\def\Ra{Ra_{\phi}}
\def\Pra{Pr}
\def\RF{R_F}
\newcommand{\avgphiz}[1]{\left < #1 \right>_{\varphi z}}

\shorttitle{Mixed convection regimes in a thin cylindrical gas layer}
\shortauthor{F. Rein, L. Car\'enini, F. Fichot, B. Favier, M. Le Bars}

\title{Experimental study of the convection in a thin cylindrical gas layer with imposed bottom and top fluxes and imposed side temperature}

\author{F. Rein \aff{1,2}
  \corresp{\email{florian.rein@protonmail.com}},
   L. Car\'enini\aff{2}, F. Fichot\aff{2}, B. Favier\aff{1}, M. Le Bars\aff{1}}

\affiliation{\aff{1}Aix Marseille Univ, CNRS, Centrale Marseille, IRPHE, Marseille, France \aff{2} IRSN, St Paul lez Durance, France}

\begin{document}

\maketitle

\begin{abstract}
We investigate convection in a thin cylindrical gas layer with an imposed flux at the bottom and a fixed temperature along the side, using a combination of Direct Numerical Simulations and laboratory experiments. 
The experimental approach allows us to extend by two orders of magnitude the explored range in terms of flux Rayleigh number. 
We identify a scaling law governing the root mean square horizontal velocity and explain it through a dimensional analysis based on heat transport in the turbulent regime. 
Using Particle Image Velocimetry, we experimentally confirm, for the most turbulent regimes, the presence of a drifting persistent pattern consisting of radial branches, as identified by \cite{reinjfm}. 
We characterize the angular drift frequency and azimuthal wave number of this pattern as functions of the Rayleigh number. 
The system exhibits a wide distribution of heat flux across various timescales, with the longest fluctuations attributed to the branch pattern and the shortest to turbulent fluctuations.
Consequently, the branch pattern must be considered to better forecast important wall heat flux fluctuations, a result of great relevance in the context of nuclear safety, the initial motivation for our study.
\end{abstract}

\begin{keywords}
 convection, laboratory experiments
\end{keywords}

\section{Introduction}
The present study follows up the numerical study of \cite{reinjfm} and is motivated by nuclear safety issues.
During a severe accident in a nuclear power plant, the radioactive fuel and metallic components of the reactor melt, forming a fluid known as corium. 
The corium relocates from the reactor core to the lower plenum of the reactor vessel, where non-miscible oxidic and metallic phases separate. 
The oxide phase  contains the majority of radioactive elements. It thus heats from below the less dense and thinner liquid metal phase floating on its surface, which then  concentrates the heat towards the vessel wall. This phenomenon is referred to as the "focusing effect" in nuclear safety literature. 
Understanding heat transfer through the top metal layer is crucial for predicting vessel failure or ensuring its integrity when In-Vessel Retention is employed as a severe accident (SA) management strategy \citep{THEOFANOUS,IVR}.

A previous study of the metal layer dynamics \citep{reinjfm}  highlighted the interplay between the Rayleigh-Bénard convection triggered by the bottom heating and the vertical convection resulting from the lateral cooling. 
Both configurations have been the subject of numerous studies, see e.g. \cite[][]{Ahlers2009,cp,doering_2002,verzicco_2008,johnston,fantuzzi_2018} and 
\cite[][]{Batchelor,churchill,bejan,GEORGE,wells_worster_2008,Ng_2015,shishkina}, respectively. 
In integral severe accident codes, like the ASTEC code developed by IRSN \citep{CHATELARD2014}, the entire process of a reactor core meltdown accident is simulated, from initiating events to the release of radioactive materials.
Different modules address different aspects of the accident, the corium behaviour in the lower plenum of the vessel being one of them \citep{Fdc-ASTEC}.
In such codes, the focusing effect evaluation is based on a simplified approach proposed by \cite{THEOFANOUS} (denoted 0D model), which combines correlations from both Rayleigh-B\'enard and natural convection to characterize the fluid by a single, mean surface temperature.
It is assumed that the fluid in the bulk is thoroughly mixed and that the vertical heat transfer is symmetrical, meaning that the temperature difference between the bottom and the bulk is equal to the temperature difference between the bulk and the top.

The 0D model was experimentally tested using a rectangular geometry and water as a simulant for the metal layer. 
Initially, assumptions regarding a well-mixed fluid and symmetry were verified through the MELAD experiment \citep{THEOFANOUS}. 
Further examination was conducted in the BALI-metal experiment, focusing on higher aspect ratios \citep{bali}.
Findings indicated that the initial model might overestimate the focusing effect for shallow layer thickness. 
Recent HELM \citep{HELM} and HELM-LR \citep{HELM-LR} experiments, employing a cylindrical geometry but maintaining water as the simulant, aimed at refining these conclusions for various aspect ratios. 
Additionally, numerical simulations of the metal layer were carried out \citep{IVMR} and revealed a significant impact of the fluid properties (water versus metal layer) on the overall phenomenology, particularly due to differences in the Prandtl number, substantially lower for the metal layer compared to water (0.1 versus 7). 
The lateral heat flux and concentration factor were found to be up to $50\%$ higher with steel compared to water under similar boundary conditions. 
The validity of results obtained with water as a simulant for the metallic layer, as well as the derived model for SA integral codes, are therefore questionable.
Note that although the Prandtl number in liquid metals is generally lower than $0.1$, for example close to 0.026 in liquid gallium \citep{Aurnou}, the present 
metal layer is composed of a mixture of zirconium, metallic uranium, and steel.
In this configuration the expected Prandtl value is in the range [0.07-0.2] \citep{pr}.
Hence, a representative intermediate value of 0.1 is considered.

\cite{reinnureth} proposed an improvement of the 0D model, incorporating a description of the radial temperature profile at the top surface based on DNS and scaling laws (called 1D model in the following). 
However, the applicability of this 1D model was only tested numerically up to a flux Rayleigh number $\Ra=10^8$ (see the exact definition \eqref{eq:NSD} below), while the system can reach up to $\Ra=10^{14}$ during severe accidents.
Even for thin metal layers (typically below $20$cm in depth), for which the focusing effect is strongest, the Rayleigh number can reach up to $\Ra=10^{10}$.

Besides, all existing models focus on the mean value of the heat flux only, and overlook possible large amplitude fluctuations on the side boundary.
\cite{reinjfm} highlighted the presence of a three-dimensional (3D) drifting thermal structure taking the form of radial hot branches, potentially playing a crucial role in those heat transfer fluctuations, but up to now unaccounted for.

This paper presents an experimental study of the convection in a thin cylindrical gas layer with a Prandtl number $\Pra =0.7$, alongside 3D Direct Numerical Simulations (DNS) of the experimental setup using both the experimental working fluid and a fluid with a Prandtl number $\Pra=0.1$ representative of the nuclear application  \citep{IVR}. The objectives are to: (i) experimentally confirm and expand the scope of scaling laws identified through our previous numerical simulations; 
(ii) experimentally identify the pattern of  drifting branches and analyze its properties depending on the input parameters; and (iii) investigate the heat flux fluctuations at the wall and quantify the influence of the drifting branch pattern on heat transfer.
The paper is divided into four sections. 
The next section delves into the experimental setup, detailing the instrumentation and the experimental protocol.
In section 3, we outline the governing equations and the numerical simulation tool employed. 
In section 4, we present both experimental and numerical results, beginning with an analysis of mean values within the system and describing scaling laws through dimensional analysis arguments. 
Additionally, we examine the angular drift velocity and the azimuthal wave number of the drifting branch pattern in relation to the Rayleigh number. 
We also investigate the distribution of heat flux fluctuations and explore the role of the drifting pattern in heat flux transport along the side. 
The concluding section discusses the implications of our findings for nuclear safety and outlines potential avenues for future research.

\section{Experimental method}

\subsection{Experimental setup : FOCUS}
\input{Figures/fig1}
The experimental setup, named FOCUS and depicted in Figure \ref{setup}, consists of an upright acrylic cylindrical enclosure with a vertical main axis and an aspect ratio $\Gamma=R/H=4$, with $R$ the tank radius and $H$ its height. 
Inside this enclosure, a gas (air or air/\ch{SF6} mixture) is heated from below and cooled along the side. 

The choice of a gas as a simulant was motivated by the Prandtl number characteristic of the metal layer in the nuclear application, which is less than 1 ($\Pra = \nu/\kappa \simeq 0.1$, with $\nu$ the kinematic viscosity and $\kappa$ the thermal diffusivity).
When $\Pra < 1$, the viscous boundary layer is nested within the thermal boundary layer.
This structure plays a major role in heat transfers \citep{reinjfm}. 
Using gases with $\Pra\simeq 0.7 <1$ ensures maintaining a similar configuration, thus preserving similar heat transfer mechanisms, without the cost and many  difficulties of using liquid metals (including for instance their opacity or the necessity to work at higher than ambient temperature).
\ch{SF6} being denser than air, it allows reaching more turbulent regimes by significantly increasing the Rayleigh number: indeed, the thermal conductivity and the dynamic viscosity of air and \ch{SF6} are mostly similar, so the Rayleigh number (see the exact definition \eqref{eq:NSD} below) scales as the gas density squared. Note that such gases have already been successfully used to study high-Rayleigh number thermal convection in the past, see e.g. \cite{Steinberg_1999,Ahlers_2009};  and more recently \cite{CIERPKA2019109841}.

In our set-up (see Figure \ref{setup}), the gas layer is heated from below through 8 silicone heating elements (\qtyproduct{450x250}{mm} each) with a combined maximum power of \qty{7.2}{kW}, resting on a cylindrical aluminum plate with a radius of $R=\qty{44}{cm}$, for better heat flux homogenization.
These heating elements are positioned on a \qty{5}{cm} thick calcium silicate insulating plate to limit heat losses and to concentrate heat into the aluminum plate. Cooling on the side is provided by a cylindrical PMMA torus with a rectangular section of \qty{11}{cm} in height and \qty{40}{mm} in width, through which water flows at an average rate of \qty{15}{L/min}. The water temperature is regulated by a thermostatic bath with a power of \qty{1.2}{kW}. Three water inlets and outlets located at the bottom and top of the torus, arranged in a staggered pattern and spaced at \ang{60} intervals, promote water mixing through natural and forced convection. The thickness of the inner wall of the torus is only \qty{4}{mm} (minimum thickness achievable through manufacturing processes), so as to ensure a temperature as constant as possible at the wall. Note that on the edges of the bottom plate, a thin \qty{3}{mm} cork plate was added to prevent direct contact between the heating elements and the bottom of the cooling torus, while holes were drilled in the insulating plate to dissipate heat.
Note also that insulated walls can introduce side wall effects, especially when gas is used \citep{Roche2001}.
In our setup however, we expect that the temperature-regulated water inside the tank, controlled by the powerful thermal bath, largely compensates for any residual heat flux.

A transparent polycarbonate lid with a thickness of \qty{35}{mm} is used to maintain the gas in the tank and to constrain the outgoing heat flux, while allowing flow visualization from above. This outgoing heat flux through the lid is a parameter of our study and accounts for approximately 50\% of the input heat flux across all experiments (see Appendix \ref{annexe}).
The choice of using a polycarbonate cover was guided by the temperature range that can be reached with the experimental setup, in order to prevent any changes in lid opacity and ensure optimal flow visualization. 

Finally, to minimize the \ch{SF6} mixing with the external environment, the entire setup is enclosed in a large square polymethyl methacrylate (PMMA) tank measuring ($\qtyproduct{1220x1220x470}{mm}$).

The operational temperature range of the device has been limited to less than \qty{100}{\degreeCelsius}, reaching a maximal bulk temperature of $64^\circ$C.
Indeed, since we use gas as a simulant for the incompressible metal layer, we want to avoid excessive temperature differences which would result in significant and undesirable compressibility effects.
For an ideal gas, the density variation can be estimated as the ratio of the largest temperature change between the initial and statistically stationary states to the mean temperature in the statistically stationary state.
Throughout all the experiments, we measured an average relative density variation of \qty{5}{\%}, peaking at \qty{11.5}{\%} for the highest Rayleigh number obtained.

\subsection{Instrumentation}
\subsubsection{Heat flux sensor}
In order to measure heat flux fluctuations through the side wall, we use a thermal flux sensor ("gSkin" manufactured by "greenTEG") measuring \qtyproduct{18x18x0.5}{mm} and positioned on the inner wall of the torus. 
It has a response time on the order of \qty{0.7}{s} and its sensitivity depends on temperature via the relation:
\begin{equation}
s=S_0 + (\theta_0-\theta_S)S_c,
\end{equation}
with $S_0=\qty{39.92}{\micro\volt\per\watt\square\meter}$, $\theta_S=\qty{22.5}{\degreeCelsius} $, $S_c=\qty{0.0499}{\micro\volt\per\watt\square\meter\per\degreeCelsius}.$
$\theta_0$ corresponds to the temperature value of the substrate on which the flux sensor is placed, hence the cooling temperature here. 
During the measurement campaign, the sensor was mainly placed at mid-height of the torus, using thermal paste to avoid any parasitic thermal resistance and to ensure good adhesion to the wall over the entire surface despite the local curvature.
\subsubsection{PT100 and thermocouples}
We also deployed a set of temperature measurements inside the setup, as sketched in Figure \ref{setup}b. Six thermocouples are positioned at mid-height of the tank, distributed along two radii offset by \ang{90}, and located at $r=\qty{10}{cm}$, $\qty{20}{cm}$, and $\qty{30}{cm}$ from the tank center. 
This specific arrangement aims at characterizing the dynamics of the thermal branch pattern, particularly by measuring its drift frequency. 
Additionally, we have inserted and leveled a PT100 sensor in the center of the aluminum bottom plate. 
Two other PT100 sensors are used to measure the inlet and outlet temperatures of the water in the torus to estimate the extracted heat flux. 
To complete the setup, two additional thermocouples are positioned above and below the lid at its center. 
All signals are acquired using National Instruments modules (NI9212, NI9211, and NI9217 modules for the acquisition of thermocouples, flux sensor, and PT100 sensors, all connected to an NI cDAQ-9174 rack). 
Those measurements contribute to a comprehensive thermal characterization of the system. In particular, they allow for quantifying the thermal losses and the distribution of heat flux between the lateral and upper surfaces, as described in Appendix \ref{energybalance}.

\subsubsection{Particle Image Velocimetry (PIV)} \label{sec:PIV}
We use time-resolved PIV to determine the horizontal velocities in an horizontal laser sheet positioned \qty{6}{cm} above the aluminum plate. This sheets is produced by a 10 Watts Asur light systems 488-532 nm CW Laser with a Powell lens.
Finding particles acting as passive scalars and persisting during a long time in our closed gas system proved particularly challenging. 
Initially, the use of fine water droplets seemed promising, but in an environment subjected to high temperatures, the lifespan of these droplets was too short due to rapid evaporation. 
This issue was resolved by adding a small fraction of UCON oil to the water droplets.  
However, a fine balance had to be struck, as a substantial addition of oil leads to an increase in the typical size of the droplets and thus to rapid sedimentation, altering their behavior as passive scalars. 
An oil concentration of \qty{3.5}{\%} by volume was identified as an optimal compromise \citep{dorel2023}, allowing the droplets to persist for a period of 30 to 40 minutes while faithfully tracking flow motions. 
These particles are produced and introduced into the system via a \qty{1}{cm} diameter conduit on the side of the cover (within its thickness) using an atomizer. 
Their average size is of the order of $10 \mu$m.
An estimate of the viscous relaxation time of the droplet motion shows that it is much smaller than the Kolmogorov time, indicating that the droplets behave as passive tracers. Furthermore, we verified that the volume of liquid introduced into the system is too small to have a significant thermal impact on the gas properties.

The droplets motion is tracked using a camera positioned above the setup (Blackfly-U3-51S5M FLIR  with a Fujinon \qty{12.5}{mm} lens). 
We narrowed the camera's field of view to achieve a sampling frequency ranging from 100 fps to 180 fps, resulting in images of \qtyproduct{1500x700}{pixels}. The observation area forms a rectangle of \qtyproduct{34x16}{cm}, offset by \qty{18}{cm} from the center (see Figure \ref{setup}a). 
The velocity fields are derived from these images using the DPIVSoft program developed by \cite{Meunier2003}. 
We consider \qtyproduct{24x24}{pixels} boxes with \qty{50}{\%} overlap on \qtyproduct{1500x700}{pixels} images, resulting in a velocity field of $125 \times 58$ vectors. 
Throughout the results, the percentage of spurious vectors in the velocity field never exceeded an average of \qty{5}{\%}.

\subsection{Experimental protocol}
We now detail the experimental protocol implemented during the measurement campaign.
We begin by initiating the acquisition system for temperature and flux.
In the case of \ch{SF6} usage, plastic sheets are placed above the setup and fixed to the outer basin to contain the gas. 
The outer basin is first filled completely with \ch{SF6} from the bottom, until gas is detected on the surface of the sheets using a gas detector. 
Then, the experimental setup is filled with \ch{SF6} while aiming to maximize density. There, we temporarily use a densimeter (WIKIA Tech, Avenisense NorthDome) placed inside the system, with its sensor positioned \qty{2}{cm} above the aluminum plate, near the edge opposite to the flux sensors.

Water circulation and water cooling via the thermostated bath are initiated. The bath temperature setpoint is chosen to maintain a  difference of $\sim$ \qty{5}{\degreeCelsius} with the ambient temperature, aiming to keep the same external heat losses between each experiment (see details in Appendix \ref{energybalance}). Finally, the bottom heating system is activated by selecting the electrical power to be supplied to the heating elements.

We now enter the transient regime. 
When using \ch{SF6}, it is expected that the density inside the setup decreases as it is not gas-tight, and \ch{SF6} diffuses into the ambient air. 
If the density drops below the desired threshold, more \ch{SF6} is injected into the system until the desired density is reached.
The gas in the system is then a mixture of air and \ch{SF6}. 
The mass fraction of \ch{SF6}, denoted $y$, is determined via a density measurement further detailed in Appendix \ref{gasmixture}.

During this transient regime, the characteristic time and thermal losses are evaluated (see Appendix \ref{energybalance}).
The steady-state regime is achieved when the heating time exceeds 3 characteristic times, i.e. in about $\sim$ 9h. 
At this stage, it must be ensured that the temporal variation of temperature is less than \qty{1}{\degreeCelsius/\hour}.

Once the steady state is reached, flux data is collected for 3 to 6 diffusive times ($\sim$ 1h).
During this period, the necessary elements for PIV measurements are prepared, including the water/UCON oil mixture to generate PIV particles. 
After the flux data acquisition is completed, the seeding phase begins.
A low-power horizontal laser sheet (\qty{0.2}{W}) is set up, an air/\ch{SF6} injection flow rate through the atomizer is fixed, and the setup is seeded for a predefined duration to control the quantity of injected PIV particles, ensuring experiment reproducibility.
The injection of these particles disturbs the flow and the thermal fluxes involved. It is necessary to wait for the particle jet to homogenize and the thermal signals to return to steady state before starting the flow video acquisition. 
The laser power, gain, sampling frequency, and exposure time are then adjusted to identify and track the movements of the PIV particles. 
Series of video acquisition are then conducted, each for a duration of approximately one minute (a duration for which we verify a constant sampling frequency).

\section{Mathematical and numerical formulation}
\subsection{Governing equations}
\begin{figure}
     \centering

  
\tikzset {_o9sekbrdd/.code = {\pgfsetadditionalshadetransform{ \pgftransformshift{\pgfpoint{0 bp } { -3.5 bp }  }  \pgftransformrotate{-269 }  \pgftransformscale{2 }  }}}
\pgfdeclarehorizontalshading{_0y226dau0}{150bp}{rgb(0bp)=(0.82,0.01,0.11);
rgb(37.5bp)=(0.82,0.01,0.11);
rgb(57.23214285714286bp)=(1,1,1);
rgb(100bp)=(1,1,1)}

  
\tikzset {_7bjqkhyjr/.code = {\pgfsetadditionalshadetransform{ \pgftransformshift{\pgfpoint{0 bp } { 0 bp }  }  \pgftransformrotate{0 }  \pgftransformscale{2.6 }  }}}
\pgfdeclarehorizontalshading{_tjvjvgiez}{150bp}{rgb(0bp)=(0.29,0.56,0.89);
rgb(37.5bp)=(0.29,0.56,0.89);
rgb(45.69940294538225bp)=(0.82,0.92,0.98);
rgb(62.5bp)=(0.82,0.01,0.11);
rgb(100bp)=(0.82,0.01,0.11)}

  
\tikzset {_avktijkzp/.code = {\pgfsetadditionalshadetransform{ \pgftransformshift{\pgfpoint{0 bp } { 0 bp }  }  \pgftransformrotate{0 }  \pgftransformscale{2.6 }  }}}
\pgfdeclarehorizontalshading{_iccg9onrs}{150bp}{rgb(0bp)=(0.82,0.01,0.11);
rgb(37.5bp)=(0.82,0.01,0.11);
rgb(53.39285714285714bp)=(0.82,0.92,0.98);
rgb(62.5bp)=(0.29,0.56,0.89);
rgb(100bp)=(0.29,0.56,0.89)}
\tikzset{every picture/.style={line width=0.75pt}} 

\begin{tikzpicture}[x=0.75pt,y=0.75pt,yscale=-1,xscale=1]

\draw  [draw opacity=0][shading=_0y226dau0,_o9sekbrdd][line width=0.75]  (330.97,12605.64) -- (641.36,12605.64) -- (641.36,12612.32) -- (330.97,12612.32) -- cycle ;
\draw  [draw opacity=0][shading=_tjvjvgiez,_7bjqkhyjr][line width=0.75]  (330.73,12612.33) -- (330.73,12532.04) -- (335.3,12532.02) -- (335.3,12612.31) -- cycle ;
\draw  [draw opacity=0][shading=_iccg9onrs,_avktijkzp][line width=0.75]  (639.84,12612.34) -- (639.84,12532.05) -- (644.41,12532.03) -- (644.41,12612.32) -- cycle ;
\draw [color={rgb, 255:red, 74; green, 74; blue, 74 }  ,draw opacity=1 ]   (490.27,12525.85) -- (646.07,12525.85) ;
\draw [shift={(648.07,12525.85)}, rotate = 180] [color={rgb, 255:red, 74; green, 74; blue, 74 }  ,draw opacity=1 ][line width=0.75]    (6.56,-2.94) .. controls (4.17,-1.38) and (1.99,-0.4) .. (0,0) .. controls (1.99,0.4) and (4.17,1.38) .. (6.56,2.94)   ;
\draw [shift={(488.27,12525.85)}, rotate = 0] [color={rgb, 255:red, 74; green, 74; blue, 74 }  ,draw opacity=1 ][line width=0.75]    (6.56,-2.94) .. controls (4.17,-1.38) and (1.99,-0.4) .. (0,0) .. controls (1.99,0.4) and (4.17,1.38) .. (6.56,2.94)   ;
\draw [color={rgb, 255:red, 74; green, 74; blue, 74 }  ,draw opacity=1 ]   (324.8,12612.42) -- (324.8,12531.68) ;
\draw [shift={(324.8,12529.68)}, rotate = 90] [color={rgb, 255:red, 74; green, 74; blue, 74 }  ,draw opacity=1 ][line width=0.75]    (6.56,-2.94) .. controls (4.17,-1.38) and (1.99,-0.4) .. (0,0) .. controls (1.99,0.4) and (4.17,1.38) .. (6.56,2.94)   ;
\draw [shift={(324.8,12614.42)}, rotate = 270] [color={rgb, 255:red, 74; green, 74; blue, 74 }  ,draw opacity=1 ][line width=0.75]    (6.56,-2.94) .. controls (4.17,-1.38) and (1.99,-0.4) .. (0,0) .. controls (1.99,0.4) and (4.17,1.38) .. (6.56,2.94)   ;
\draw    (299.4,12609.22) -- (306.81,12609.22) ;
\draw [shift={(296.4,12609.22)}, rotate = 0] [fill={rgb, 255:red, 0; green, 0; blue, 0 }  ][line width=0.08]  [draw opacity=0] (3.57,-1.72) -- (0,0) -- (3.57,1.72) -- cycle    ;
\draw    (306.4,12609.22) -- (306.4,12599.57) ;
\draw [shift={(306.4,12596.57)}, rotate = 90] [fill={rgb, 255:red, 0; green, 0; blue, 0 }  ][line width=0.08]  [draw opacity=0] (3.57,-1.72) -- (0,0) -- (3.57,1.72) -- cycle    ;

\draw [color={rgb, 255:red, 208; green, 2; blue, 27 }  ,draw opacity=1 ]   (481.98,12553.06) .. controls (480.3,12551.4) and (480.29,12549.73) .. (481.95,12548.06) .. controls (483.6,12546.39) and (483.59,12544.72) .. (481.92,12543.06) -- (481.91,12542.36) -- (481.86,12534.36) ;
\draw [shift={(481.85,12532.36)}, rotate = 89.64] [color={rgb, 255:red, 208; green, 2; blue, 27 }  ,draw opacity=1 ][line width=0.75]    (7.65,-2.3) .. controls (4.86,-0.97) and (2.31,-0.21) .. (0,0) .. controls (2.31,0.21) and (4.86,0.98) .. (7.65,2.3)   ;
\draw [color={rgb, 255:red, 74; green, 74; blue, 74 }  ,draw opacity=1 ]   (393.42,12551.18) -- (408.66,12534.26) ;
\draw [shift={(409.99,12532.77)}, rotate = 132] [color={rgb, 255:red, 74; green, 74; blue, 74 }  ,draw opacity=1 ][line width=0.75]    (7.65,-2.3) .. controls (4.86,-0.97) and (2.31,-0.21) .. (0,0) .. controls (2.31,0.21) and (4.86,0.98) .. (7.65,2.3)   ;
\draw [color={rgb, 255:red, 74; green, 74; blue, 74 }  ,draw opacity=1 ]   (609.39,12592.26) -- (609.39,12605.4) ;
\draw [shift={(609.39,12607.4)}, rotate = 270] [color={rgb, 255:red, 74; green, 74; blue, 74 }  ,draw opacity=1 ][line width=0.75]    (7.65,-2.3) .. controls (4.86,-0.97) and (2.31,-0.21) .. (0,0) .. controls (2.31,0.21) and (4.86,0.98) .. (7.65,2.3)   ;
\draw    (487.57,12607.64) -- (487.57,12616.35) ;
\draw [color={rgb, 255:red, 208; green, 2; blue, 27 }  ,draw opacity=1 ]   (493.1,12553.06) .. controls (491.42,12551.4) and (491.41,12549.73) .. (493.07,12548.06) .. controls (494.72,12546.39) and (494.71,12544.72) .. (493.04,12543.06) -- (493.03,12542.36) -- (492.98,12534.36) ;
\draw [shift={(492.97,12532.36)}, rotate = 89.64] [color={rgb, 255:red, 208; green, 2; blue, 27 }  ,draw opacity=1 ][line width=0.75]    (7.65,-2.3) .. controls (4.86,-0.97) and (2.31,-0.21) .. (0,0) .. controls (2.31,0.21) and (4.86,0.98) .. (7.65,2.3)   ;
\draw [color={rgb, 255:red, 208; green, 2; blue, 27 }  ,draw opacity=1 ]   (472.47,12638.39) .. controls (470.79,12636.73) and (470.78,12635.06) .. (472.44,12633.39) .. controls (474.09,12631.72) and (474.08,12630.05) .. (472.41,12628.39) -- (472.4,12627.69) -- (472.35,12619.69) ;
\draw [shift={(472.34,12617.69)}, rotate = 89.64] [color={rgb, 255:red, 208; green, 2; blue, 27 }  ,draw opacity=1 ][line width=0.75]    (7.65,-2.3) .. controls (4.86,-0.97) and (2.31,-0.21) .. (0,0) .. controls (2.31,0.21) and (4.86,0.98) .. (7.65,2.3)   ;
\draw [color={rgb, 255:red, 208; green, 2; blue, 27 }  ,draw opacity=1 ]   (483.13,12638.39) .. controls (481.45,12636.73) and (481.44,12635.06) .. (483.1,12633.39) .. controls (484.75,12631.72) and (484.74,12630.05) .. (483.07,12628.39) -- (483.06,12627.69) -- (483.01,12619.69) ;
\draw [shift={(483,12617.69)}, rotate = 89.64] [color={rgb, 255:red, 208; green, 2; blue, 27 }  ,draw opacity=1 ][line width=0.75]    (7.65,-2.3) .. controls (4.86,-0.97) and (2.31,-0.21) .. (0,0) .. controls (2.31,0.21) and (4.86,0.98) .. (7.65,2.3)   ;
\draw [color={rgb, 255:red, 208; green, 2; blue, 27 }  ,draw opacity=1 ]   (492.27,12638.39) .. controls (490.59,12636.74) and (490.58,12635.07) .. (492.23,12633.39) .. controls (493.88,12631.72) and (493.87,12630.05) .. (492.2,12628.39) -- (492.2,12627.69) -- (492.15,12619.69) ;
\draw [shift={(492.14,12617.69)}, rotate = 89.64] [color={rgb, 255:red, 208; green, 2; blue, 27 }  ,draw opacity=1 ][line width=0.75]    (7.65,-2.3) .. controls (4.86,-0.97) and (2.31,-0.21) .. (0,0) .. controls (2.31,0.21) and (4.86,0.98) .. (7.65,2.3)   ;
\draw [color={rgb, 255:red, 208; green, 2; blue, 27 }  ,draw opacity=1 ]   (501.27,12639.1) .. controls (499.6,12637.45) and (499.59,12635.78) .. (501.24,12634.1) .. controls (502.89,12632.43) and (502.88,12630.76) .. (501.21,12629.1) -- (501.21,12628.4) -- (501.16,12620.4) ;
\draw [shift={(501.14,12618.4)}, rotate = 89.64] [color={rgb, 255:red, 208; green, 2; blue, 27 }  ,draw opacity=1 ][line width=0.75]    (7.65,-2.3) .. controls (4.86,-0.97) and (2.31,-0.21) .. (0,0) .. controls (2.31,0.21) and (4.86,0.98) .. (7.65,2.3)   ;
\draw [color={rgb, 255:red, 74; green, 74; blue, 74 }  ,draw opacity=1 ]   (621.57,12578.87) -- (636.32,12578.87) ;
\draw [shift={(638.32,12578.87)}, rotate = 180] [color={rgb, 255:red, 74; green, 74; blue, 74 }  ,draw opacity=1 ][line width=0.75]    (7.65,-2.3) .. controls (4.86,-0.97) and (2.31,-0.21) .. (0,0) .. controls (2.31,0.21) and (4.86,0.98) .. (7.65,2.3)   ;
\draw [color={rgb, 255:red, 74; green, 144; blue, 226 }  ,draw opacity=1 ][line width=0.75]    (330.73,12532.03) -- (330.73,12612.31) ;
\draw  [line width=0.75]  (330.73,12532.04) -- (644.41,12532.04) -- (644.41,12612.33) -- (330.73,12612.33) -- cycle ;
\draw [color={rgb, 255:red, 74; green, 144; blue, 226 }  ,draw opacity=1 ][line width=0.75]    (644.41,12532.03) -- (644.41,12612.31) ;
\draw [color={rgb, 255:red, 74; green, 144; blue, 226 }  ,draw opacity=1 ][line width=0.75]    (330.73,12532.05) -- (330.73,12612.33) ;

\draw (575,12575) node [anchor=north west][inner sep=0.75pt]   [align=left] {{\scriptsize \textcolor[rgb]{0.29,0.29,0.29}{No slip}}};
\draw (349,12555) node [anchor=north west][inner sep=0.75pt]   [align=left] {{\scriptsize \textcolor[rgb]{0.29,0.29,0.29}{Free slip / No slip}}};
\draw (477.09,12638.47) node [anchor=north west][inner sep=0.75pt]  [font=\footnotesize]  {$\textcolor[rgb]{0.82,0.01,0.11}{\phi }\textcolor[rgb]{0.82,0.01,0.11}{_{\text{in}}}$};
\draw (484.14,12592.95) node [anchor=north west][inner sep=0.75pt]  [font=\tiny]  {$0$};
\draw (455.97,12551.48) node [anchor=north west][inner sep=0.75pt]  [font=\footnotesize]  {$\textcolor[rgb]{0.82,0.01,0.11}{\phi }\textcolor[rgb]{0.82,0.01,0.11}{_{\text{out}}}\textcolor[rgb]{0.82,0.01,0.11}{=\phi }\textcolor[rgb]{0.82,0.01,0.11}{_{\text{in}}}\textcolor[rgb]{0.82,0.01,0.11}{\ R}\textcolor[rgb]{0.82,0.01,0.11}{_{F}}$};
\draw (646,12561) node [anchor=north west][inner sep=0.75pt]  [font=\footnotesize,color={rgb, 255:red, 74; green, 144; blue, 226 }  ,opacity=1 ]  {$\theta _{0}$};
\draw (300.15,12562.83) node [anchor=north west][inner sep=0.75pt]  [font=\footnotesize]  {$H$};
\draw (565.43,12504) node [anchor=north west][inner sep=0.75pt]  [font=\footnotesize]  {$R$};
\draw (294,12584.07) node [anchor=north west][inner sep=0.75pt]  [font=\tiny]  {\normalsize{$\mathbf{e_{z}}$}};
\draw (285,12597) node [anchor=north west][inner sep=0.75pt]  [font=\tiny]  {\normalsize{$\mathbf{e_{r}}$}};

\end{tikzpicture}

     \caption{Sketch of the modeled fluid layer in a vertical plane through the cylinder.
     The top boundary condition is no-slip for the gas experiment simulations, representing the lid used in the setup, while a free-slip condition is applied to model the metal layer in the nuclear application.}
    \label{intro}
\end{figure}
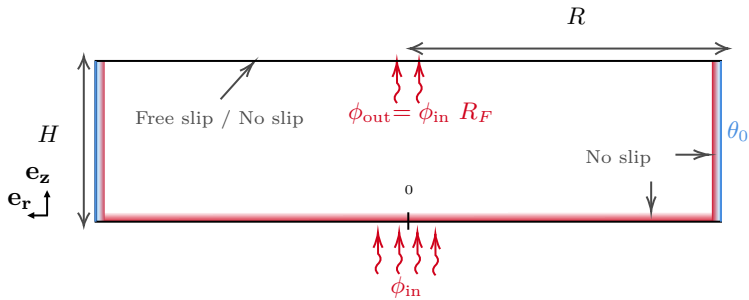

Following \cite{reinjfm}, we model the experimental setup considering an incompressible fluid in the Boussinesq approximation, confined in a cylinder of thickness $H$ and radius $R$, with gravity directed downward (see Figure \ref{intro}). 
Heating from below is applied with a uniform heat flux $\phi_{\textrm{in}}$, while cooling from above occurs with a uniform outgoing flux $\phi_{\textrm{out}}$.
Notice that this upper boundary condition is an approximation, since the experimental outgoing flux is dynamically constrained by the convective heat transfers within the system and may change radially. Nonetheless, the thermal budget described in Appendix \ref{energybalance} confirms that the surface average measurement and the local, center plate measurement of the outgoing flux have close values, hence validating this assumption with an outgoing flux $\phi_{\textrm{out}}$ ranging between $40-60\%$ of the incoming one. 
The study hence focuses on cases where $\phi_{\textrm{in}}$ and $\phi_{\textrm{out}}$ are unequal, resulting in residual heat flux escaping through the side boundary, whose temperature is fixed at $\theta_0$. 
No-slip conditions are applied to the bottom and side boundaries, while two configurations are considered for the top boundary in the DNS: a no-slip condition to represent the experimental lid, and a free-slip condition to model the liquid metal layer in the nuclear application.

Lengths are rescaled using the height of the cylinder $H$, while time is rescaled using the vertical diffusive timescale $H^2/\kappa$. 
In our set-up, the two main control parameters are the bottom heat flux and the side temperature, which is the only imposed temperature. Hence, 
the non-dimensional temperature $T$ is defined relative to the imposed side temperature and is rescaled using the imposed bottom flux $\phi_{\textrm{in}}$,

\begin{equation}
\label{ndt}
    T=\frac{k}{\phi_{\textrm{in}}H}\left(\theta-\theta_0\right) \ ,
\end{equation}
where $k$ is the thermal conductivity.
The non-dimensional conservation equations of momentum, mass and energy are then
\begin{equation}
\begin{aligned}
\frac{1}{Pr}\left(\frac{\partial\bm{u}}{\partial t}+ \bm{u\cdot\nabla}\bm{u}\right) =-\bm{\nabla}P +Ra_{\phi} T\bm{e}_z+\bm{\nabla}^2\bm{u} \ ,
\end{aligned}
\label{qdm}
\end{equation}
\begin{equation}
\bm{\nabla\cdot u}=0 \ ,
\label{m}
\end{equation}
\begin{equation}
\frac{\partial T}{\partial t}+ \bm{u}\cdot\bm{\nabla} T=\nabla^2 T \ .
\label{nrj}
\end{equation}
$\bm{u}$ and $P$ are the non-dimensional velocity and pressure (including the hydrostatic contribution) fields.
The problem is characterized by four non-dimensional parameters:
the aspect ratio $\Gamma$, the flux ratio $R_F$, the Rayleigh number $Ra_{\phi}$ based on the heat flux imposed at the bottom $\phi_{\textrm{in}}$, the Prandtl number $Pr$ set to $0.1$ to mimic the properties of the metal layer or $0.7$ to replicate those of the gas used in the experiment. They are defined by 
\begin{equation}
\Gamma = \frac{R}{H},~~~R_F=\frac{\phi_{\textrm{out}}}{\phi_{\textrm{in}}},~~~Ra_{\phi}=\frac{\beta g \phi_{\textrm{in}} H^4}{k \nu \kappa},~~~Pr=\frac{\nu}{\kappa}=\left\{\begin{array}{ll}
        0.1 & \mbox{Application}\\
        0.7 & \mbox{Experiment}
    \end{array}
\right. \ \label{eq:NSD}
\end{equation}
where $\beta$ is the thermal expansion coefficient, assumed to be constant.
The non-dimensional boundary conditions can be written as

\begin{equation}
\begin{aligned}
\bm{u}(z=0)=\bm{0}~~~\mbox{and}~~~\frac{\partial T}{\partial z}\bigg|_{z=0}=-1 \ ,\\
\bm{u}(r=\Gamma)=\bm{0}~~~\mbox{and}~~~ T(r=\Gamma)=0 \ ,\\
\left\{\begin{array}{ll}
        \mbox{Application} : & \frac{\partial u}{\partial z}\big|_{z=1}=\frac{\partial v}{\partial z}\big|_{z=1}=w(z=1)=0\\
        \mbox{Experiment :} & \bm{u}(z=1)=\bm{0}
    \end{array}
\right. \ \mbox{and}~~~\frac{\partial T}{\partial z}\bigg|_{z=1}=-R_F,
\end{aligned}
\label{bcn}
\end{equation}
where $\bm{u}=\left(u,v,w\right)$ are the velocity components in cylindrical coordinates along the unit vectors $\bm{e}_r,\bm{e}_{\varphi},\bm{e}_z$.
\subsection{Numerical approach}\label{subsecNEK}
The governing equations \eqref{qdm}-\eqref{nrj} with corresponding boundary conditions \eqref{bcn} are computationally solved using the spectral element code \href{https://nek5000.mcs.anl.gov/}{Nek5000} \citep{Fischer1997,Deville2002}. 
The cylindrical geometry is discretized with up to $\mathcal{E}=9984$ hexahedral elements, refined near all boundaries to accurately resolve viscous and thermal boundary layers,
with typically 15 grid points across the viscous boundary layers, reaching a minimum of 10 grid points for the highest Rayleigh number.

Numerical simulations are initiated with a quiescent fluid and a uniform temperature field $T=0$ throughout the domain. 
Small temperature perturbations of amplitude $10^{-3}$ are introduced, leading to thermal convection growth during a transient phase lasting $\sim$ $5$ vertical diffusive times, with longer durations observed for higher aspect ratios ($\Gamma$). 
Once the system reaches a statistically-stationary state, various spatio-temporal averages are computed. 
Temporal and volume average operators $\left<.\right>$ are defined over the entire fluid domain volume $V$ and time $\tau$ as, 
\begin{equation}
\left<.\right>=\frac{1}{\tau V}\int_{t_0}^{t_0+\tau}\int_{V}.\mathrm{d}V\mathrm{d}t ,\
\label{avg}
\end{equation}
where $\tau$ typically ranges between $2$ and $0.1$ diffusive times for the lowest and highest Rayleigh numbers, respectively.
Additionally, adding specific variables as a subscript means that an average along those specific directions is made.
We always consider temporal averages during the statistically steady state so that the time variable is never explicitly written.
For instance, $\left<.\right>_{\varphi}$ indicates an average in time and along the azimuthal direction only.

Although most results are derived from Direct Numerical Simulations (DNS), some extreme cases necessitated filtered simulations, as described in \cite{fischer}. 
A viscous dissipation criterion was employed to distinguish between DNS and filtered simulations. 
For further details, readers are directed to \cite{reinjfm}.

\section{Experimental and numerical results}
Detailed information for all experiments and simulations is available in tables \ref{table:1expe} and  \ref{table:1num} of Appendix \ref{annexe}, respectively.
The experimental data set comprises 12 independent experiments, including 3 with \ch{SF6} mixture where the average density ranged from \qty{2.9}{kg/m^3} to \qty{4.1}{kg/m^3}.
With the heating power varied from 75 to \qty{400}{W}, we have explored the range $Ra_\phi \in [3.2\times 10^7: 7.6\times 10^9]$. 
This experimental data set is completed by 8 3D DNS of the experimental setup (with no slip condition on the upper surface and $\Pra = 0.7$), exploring the range $Ra_\phi \in [10^5: 10^8]$. 
We also performed 70 DNS with free slip condition on the upper surface and $Pr=0.1$, corresponding to the nuclear application, whose results have already been reported in \cite{reinjfm}.

\subsection{Scalings}
\begin{figure}
    \centering
    \includegraphics[width=1\linewidth]{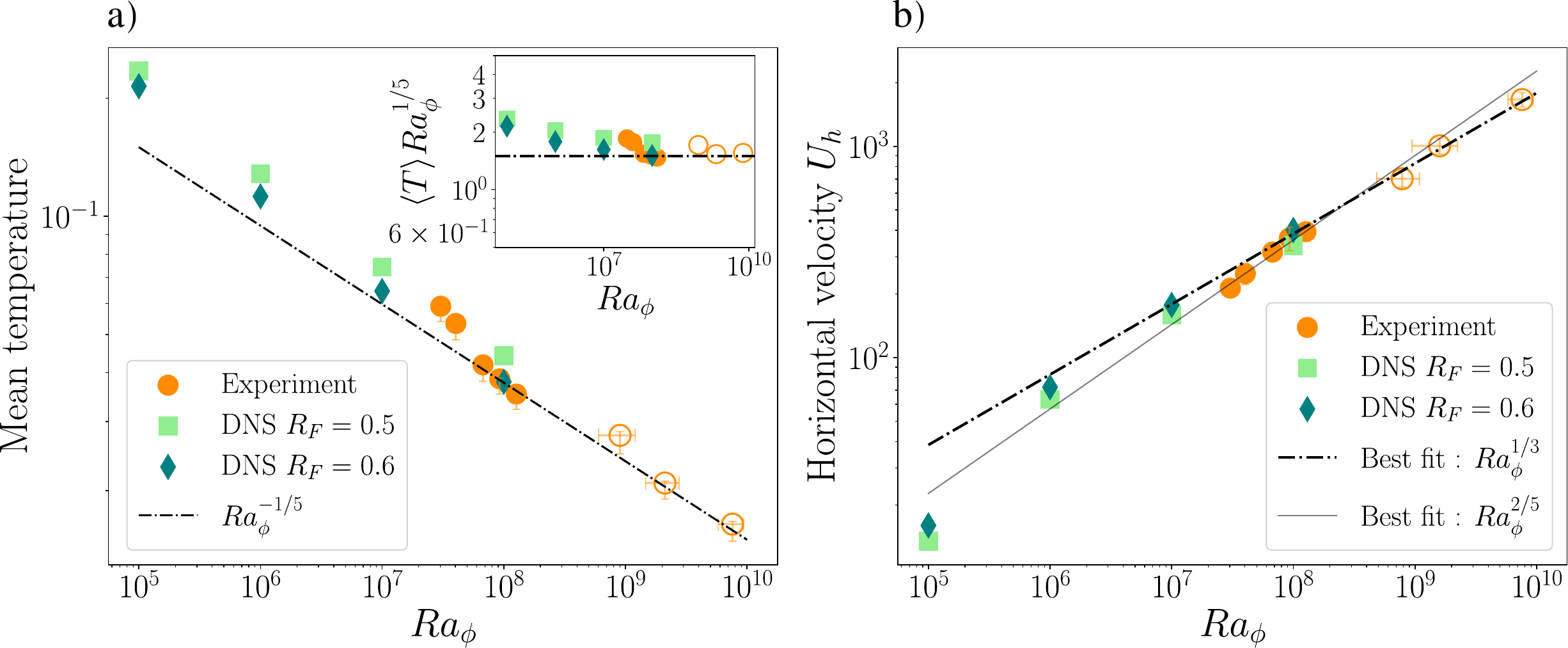}
   \caption{$a)$ Non-dimensional temperature averaged in time and over the 6 thermocouples in the bulk of the experiment, as a function of $\Ra$. The dashdot line shows the $Ra_{\phi}^{-1/5}$ asymptotic scaling from \cite{reinjfm} (the inset shows the compensated plot), $b)$ Non-dimensional horizontal RMS velocity in the $(x,y)$ plane at height $z=\qty{6}{cm}$ as a function of $\Ra$. The dashdot line and continious line show a $1/3$ and $2/5$ best fit power law based on experimental data.  
   For both plots, square ($\square$), diamond ($\Diamond$) and circle ($\circ$) symbols represent respectively data from DNS simulations at $\RF=0.5$ and $\RF=0.6$, and the experiments. Empty circles indicate when \ch{SF6} gas is used. For all DNS simulations, $\Gamma=4$ and $Pr=0.7$.}
    \label{velocityandTresult}
\end{figure}

In Figure \ref{velocityandTresult}a, the non-dimensional mean temperature is plotted against the Rayleigh number.
To do so, we averaged in time each thermocouple signal and then averaged over the 6 thermocouples within the system.
Using DNS, we compute the mean temperature following the same set of discrete locations as in the experiments, considering both $\RF=0.5$ and $\RF=0.6$ (which encompasses most of the experimental values for the flux ratio, see Table~\ref{table:1expe}), and spanning a range of $\Ra$ from $10^5$ to $10^8$.

Overall, a satisfactory agreement between DNS and experimental measurements is observed. 
One can notice here the slight influence of the flux ratio. Indeed, the two lowest experimental measurement points in terms of $\Ra$ (close to $\Ra \sim 10^7$) correspond to flux ratios $\RF\simeq 0.5$ (see Appendix \ref{energybalance}) and align well with the numerical trend observed for $\RF=0.5$. 
Similarly, the air experimental measurements near $\Ra \sim 10^8$, associated with flux ratios close to $0.6$,  show excellent agreement with the numerical trend at $\RF=0.6$.
The experimental measurements conducted with \ch{SF6} have a ratio close to $0.6$, and extend the numerical trend by nearly two orders of magnitude in terms of $\Ra$.

Also shown is the slope of the asymptotic scaling in $Ra_{\phi}^{-1/5}$ predicted by \cite{reinjfm} and which corresponds to the classical scaling of vertical convection predicted by \cite{Batchelor}; the inset graph shows the results compensated by this scaling with a good agreement above $\Ra =10^8$.
The scaling, as detailed in \cite{reinjfm}, comes from the balance at statistically stationary state between the top / bottom flux mismatch, and the lateral outgoing diffusive flux through the thermal boundary layer at the vertical boundary. As the Rayleigh number increases, the system becomes more turbulent leading to more efficient bulk mixing limited by diffusive boundary layer, hence better agreement with the scaling law.

In Figure \ref{velocityandTresult}b), we plot the norm of the non-dimensional horizontal velocity value $U_h$ as a function of the Rayleigh number for numerical simulations at $\RF =0.5$ and $0.6$,  and the  experimental data. This velocity is computed as follows:
\begin{equation}
    U_h=\left[\left<u^2\right>_{r \varphi }+\left<v^2\right>_{r \varphi }\right]^{1/2}.
\end{equation}
The averaging procedure along the azimuthal and radial directions involves the same spatial domain for the experiments and DNS (i.e. the PIV field of view, see section \ref{sec:PIV}). The time averaging is processed during a characteristic time of \qty{1}{min} for the experiments (0.1 diffusive time for air, 0.02 for \ch{SF6})  and over 0.1 diffusive time for DNS.
Each point represents a measurement of the characteristic horizontal fluid velocity localized in the PIV measurement zone.
First, it is worth noting the excellent agreement between the experimental measurements and DNS over the range of Rayleigh numbers common to both approaches. 
In particular, the experimental points between $\Ra=10^7$ and $\Ra=10^8$, conducted in air, compare favorably with the numerical trend obtained at $\Ra=10^8$. 
Additionally, the experimental measurements involving the use of \ch{SF6} extend the trend over almost 2 orders of magnitude in $\Ra$.
A power law relationship compatible with $\Ra^{1/3}$ is observed, with a best fit estimate of $\Ra^{0.36}$ using the least squares method based on the experimental points. 
Notice that in the numerical simulations, the RMS velocity exhibits minimal dependence on the flux ratio. 
This observation aligns with our previous findings using $\Pra=0.1$ and free-slip boundary conditions applied to the upper surface (not shown). In the following section, we use dimensional analysis to gain insight into the physics underlying this scaling law.

\subsection{Dimensional analysis}\label{theo}
\begin{figure}
   \centering
    \includegraphics[width=1\linewidth]{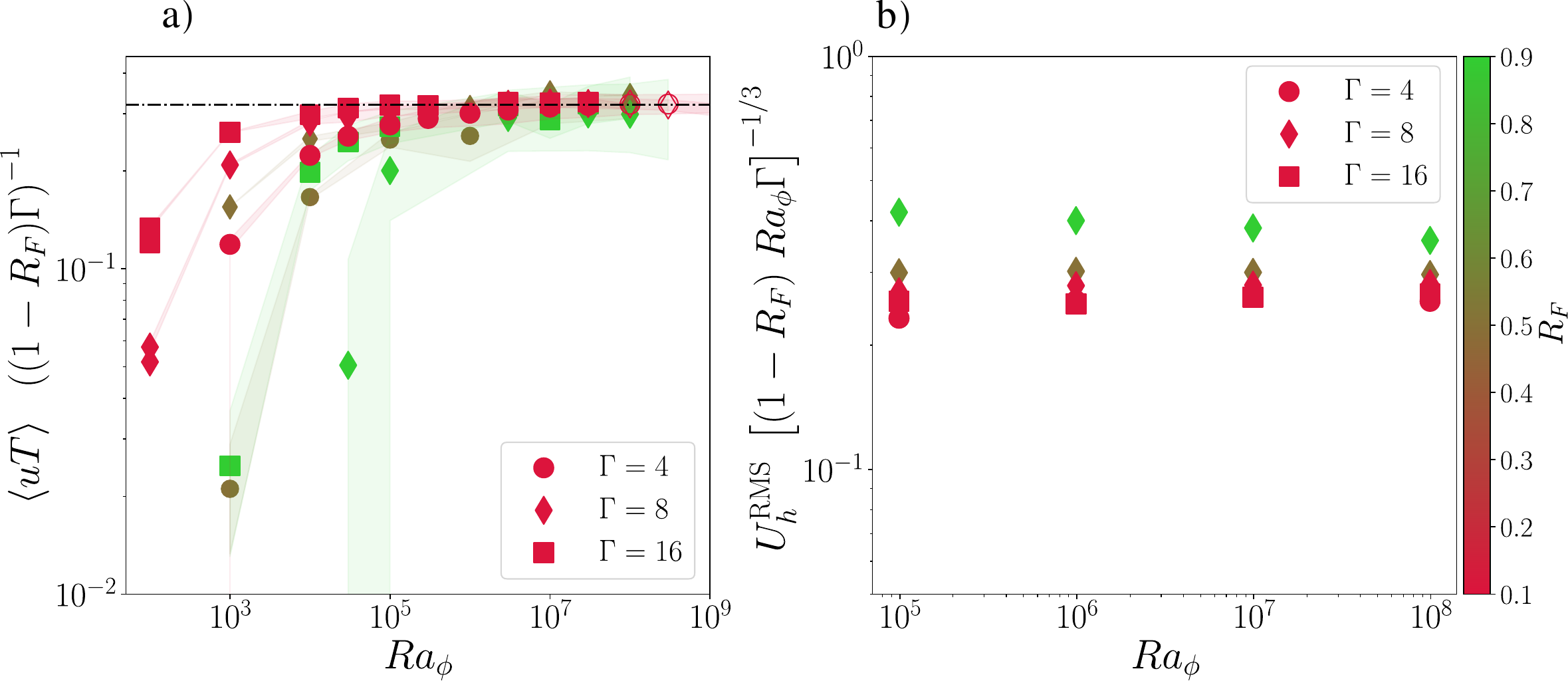}
   \caption{$a)$ Compensated horizontal advective flux following \eqref{uT} as a function of $\Ra$.
   $b)$ Compensated radial RMS velocity following \eqref{eqURMS} as a function of $\Ra$.
   For both plots, different symbols correspond to different aspect ratios $\Gamma$ and color variation from red to green involves $R_F$ variation from 0.1 to 0.9. 
   Data are from DNS with $Pr=0.1$ and a free-slip upper boundary, for which our data set is more complete in terms of $\Gamma$ and $\RF$ systematics, but the same behaviour is expected with the experimental configuration. Empty/full symbols indicate respectively filtered/DNS simulations. The shadings represent 3 times the standard deviation of the  horizontal advective flux time series at the statistically stationary state.}
   \label{URMS_scaling}
\end{figure}

Let us focus on the heat transport by considering the following non-dimensional energy equation derived from \eqref{nrj} using cylindrical coordinates : 
\begin{equation}
\frac{\partial T}{\partial t} +
    \frac{1}{r}\frac{\partial}{\partial r}\left[r \left(uT - \frac{\partial T}{\partial r}\right)\right] + 
    \frac{1}{r}\frac{\partial}{\partial \varphi}\left[vT - \frac{1}{r}\frac{\partial T}{\partial \varphi}\right]+
    \frac{\partial}{\partial z}\left[wT - \frac{\partial T}{\partial z}\right]=0 \ .
    \label{heateq}
\end{equation}
The horizontal transport of the heat flux at a radius $r$ is expressed by averaging the energy equation \eqref{heateq} in time, vertical and azimuthal directions, which leads to 
\begin{equation}
    \avgphiz{uT} - \avgphiz{\frac{\partial T}{\partial r}} =\frac{(1-R_F)}{2} r \ .
    \label{flux}
\end{equation}
This equation illustrates that the difference in heat flux between the top and bottom (i.e. the right-hand side of equation~\eqref{flux}) is transported horizontally through a combination of diffusion ($\avgphiz{\partial_r T}$) and advection ($\avgphiz{uT}$) processes. 
If we now assume the system to be fully turbulent, the radial diffusive gradient matters only within the thermal boundary layer. 
Within the bulk, we expect the balance $\avgphiz{uT} \sim \frac{(1-R_F)}{2} r$.
By averaging along the radial direction, the horizontal advective flux scales as follows:
\begin{equation}
    \langle uT \rangle = \frac{1}{\Gamma}\int^{\Gamma}_0 \avgphiz{uT} \mathrm{d}r \sim (1-R_F)\Gamma.
    \label{uT}
\end{equation}
This scaling is confirmed in figure \ref{URMS_scaling}a), where we use DNS data with $\Gamma$ ranging from $4$ to 16, $R_F$ from 0.1 to 0.9, and $Pr=0.1$. The higher the Rayleigh number, the better the agreement with the scaling given by equation~\eqref{uT}.
In particular, above $\Ra \sim 10^5$, $\langle uT \rangle$ becomes independent of the Rayleigh number and all the data collapse on the same straight line.

We now seek to explain the scaling of the horizontal velocity. 
Following the determination of the free fall velocity scaling in classical Rayleigh-Bénard convection, we assume that the dimensional horizontal velocity scale is independent of viscosity and diffusivity, but depends on its driving buoyancy, i.e. the horizontal turbulent buoyancy flux $\beta g F_h \sim \beta g \phi/\rho c_p$, related to $\langle uT \rangle$ in non-dimensional form.

Scaling analysis gives $\hat{U}^{\mathrm{RMS}}_h\sim (\beta g F_h H)^{1/3}$, where $\hat{.}$ denotes dimensional variable. 
In non-dimensional form, this becomes ${U}^{\mathrm{RMS}}_h\sim (\Ra \Pr \langle uT \rangle)^{1/3}$. Hence with (\ref{uT}),
\begin{equation}
    {U}^{\mathrm{RMS}}_h\sim (\Ra \Pr (1-R_F)\Gamma)^{1/3}.\label{eqURMS}
\end{equation}
We recover the scaling law in $\Ra^{1/3}$ identified in Figure \ref{velocityandTresult}b. We further validate this scaling law in Figure 
 \ref{URMS_scaling}b, which shows the compensated RMS horizontal velocity over a large range of $(\Gamma, \RF, \Ra)$ at $\Pra=0.1$.
From our DNS data set, we consider the inner part of the cylinder ($r\leq\Gamma/2$) in order to focus on the turbulent bulk, and we compute 
\begin{equation}
    U_{h}^{\mathrm{RMS}}=\left<\left(u - \langle u\rangle_{\varphi z r_f}\right)^2 + \left(v - \langle v\rangle_{\varphi z r_f}\right)^2 \right>_{\varphi z r_f}^{1/2},
\end{equation}
where the $r_f$ subscript indicates an average along the radial direction over $r\in [0,\Gamma/2]$.
The agreement with \eqref{eqURMS} is very good with all the data converging at large $\Ra$.
This is particularly clear for the lower flux ratios. As the value of $R_F$ increases, the relevance of a characteristic horizontal velocity derived from horizontal heat transport becomes less meaningful.
In the asymptotic case $R_F=1$, this typical velocity scale  effectively disappears since all the injected heat flux is evacuated at the top.

Note that in the context of vertical convection (flow between two differentially heated vertical walls), similar dimensional arguments have been put forward in the seminal work of \cite{GEORGE}. This scaling has been notably revisited and corroborated by subsequent studies such as \cite{VERSTEEGH1999,HÖLLING_HERWIG_2005,NG2013}.
Note also that even if we observe the same $1/3$ exponent in $\Ra$ as for horizontal convection with imposed flux conditions \citep{mullarney_2004}, the velocity corresponds in this case to the horizontal velocity within the thermal boundary layers \citep{ROSSBY,Griffiths_Hughes_Gayen_2013}, whereas in our case, it corresponds to the bulk velocity driven by the imposed buoyancy flux. 
Finally, we pointed out in our previous paper that our system shares similarity with \cite{GL} regime \textit{I-l} \citep{reinjfm}. 
If so, one might expect a typical rms velocity scaling like $\Ra^{2/5}$ (i.e. the imposed flux version of the classical $Ra^{1/2}$ scaling in the imposed temperature convection).
This does not appear to be consistent with our data, as show in Figure \ref{velocityandTresult}b). 
Presumably, the experimental range of flux ratio observed ($R_F \in [0.4 , 0.6]$) does not reach the asymptotically high value necessary to manifest the fully developed \textit{I-l} regime described by \cite{GL}, which would require larger $\Ra$ and/or $\RF$.

\subsection{Persistent drifting pattern in the turbulent regime}
\subsubsection{Characteristics}

In \cite{reinjfm}, a drifting pattern was observed at high Rayleigh numbers for low flux ratio (see their Figure 2(c) for example), but no detail was given about its properties.
During the experimental campaign discussed here, the system also showed the spontaneous emergence of such a branch pattern.
A video of experiment $3$ from table \ref{table:1expe} conducted in air at $\Ra \sim 10^7$ can be downloaded as a supplement material via this \href{https://www.dropbox.com/scl/fi/ldsco1gn0xm1ezdrm5yr9/air_slide.mp4?rlkey=8wqo2v1mwaol0zcpkj08yb69t&st=tx1apdxy&dl=1}{link}. 
In this video, we have removed the average image spanning the entire duration of the footage and applied a moving average over 5 frames across all images. 
\begin{figure}
    \centering
    \includegraphics[scale=0.4]{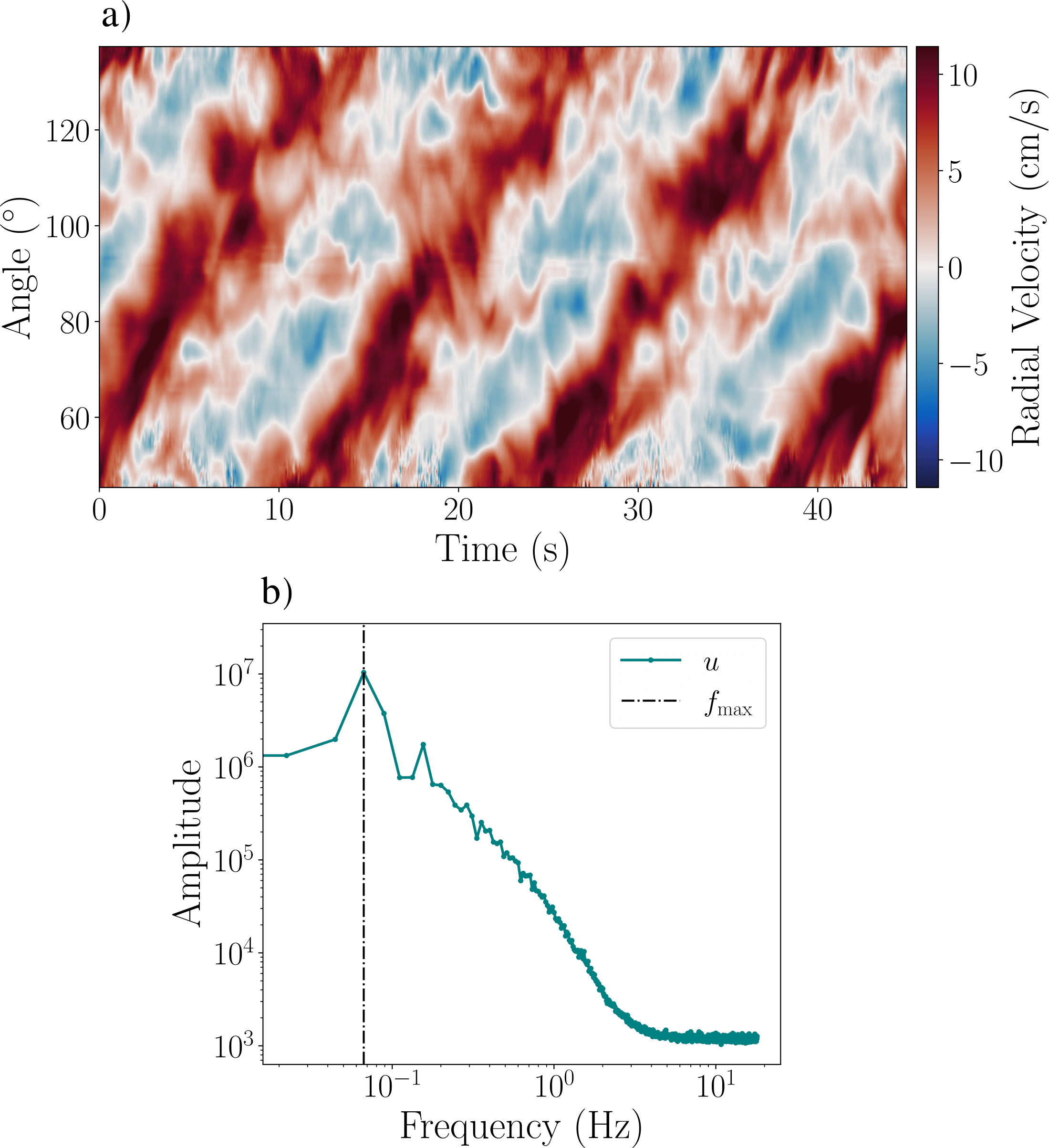}
   \caption{$a)  $Hovmöller (space-time) diagram of the radial velocity averaged along the radial direction. The radial velocity is measured at $z=6$cm (i.e, $z/H=0.54$).
   $b)$ Average of the radial velocity time spectra from each PIV measurement location. The vertical dashdot line shows the frequency ($f_{\mathrm{max}} =6.7\times10^{-2}$Hz) of the peak amplitude. For both, data correspond to experiment $9$ from table \ref{table:1expe} where $\Ra = 1.26\times10^8$, using air. 
   }
    \label{hovmoller}
\end{figure}

This adjustment allows for a clearer visualization of the movement of the PIV particles, represented by white streaks.
The video is in real-time, except for the clear appearance of a branch where a slowdown was applied.
It shows a branch of the pattern, visible as extended white streaks moving from right to left.

The drifting branches are also visible through the velocity fields obtained from PIV processing, as for instance on the Hovmöller diagram shown in Figure \ref{hovmoller}.
Thick bands associated with the highest radial velocities indicate the passage of the branches in the camera's field of view. The global angular drift frequency of the whole pattern $\omega_p$ can be deduced
from such diagrams by tracking a branch of the pattern. The observed drift can occur in both clockwise and counter-clockwise directions in independent realisations of the same configuration. This indicates that there is no preferred rotation direction as expected from the problem axisymmetry.
A local frequency $\omega$ can also be computed from the temporal spectrum of the radial velocity at each spatial location.
The spectra are then averaged over the PIV measurement area for experiments and over the whole domain for DNS.
An exemple based on the PIV measurment from experiment 9 is shown \autoref{hovmoller}b).

\begin{figure}
    \centering
    \includegraphics[width=1\linewidth]{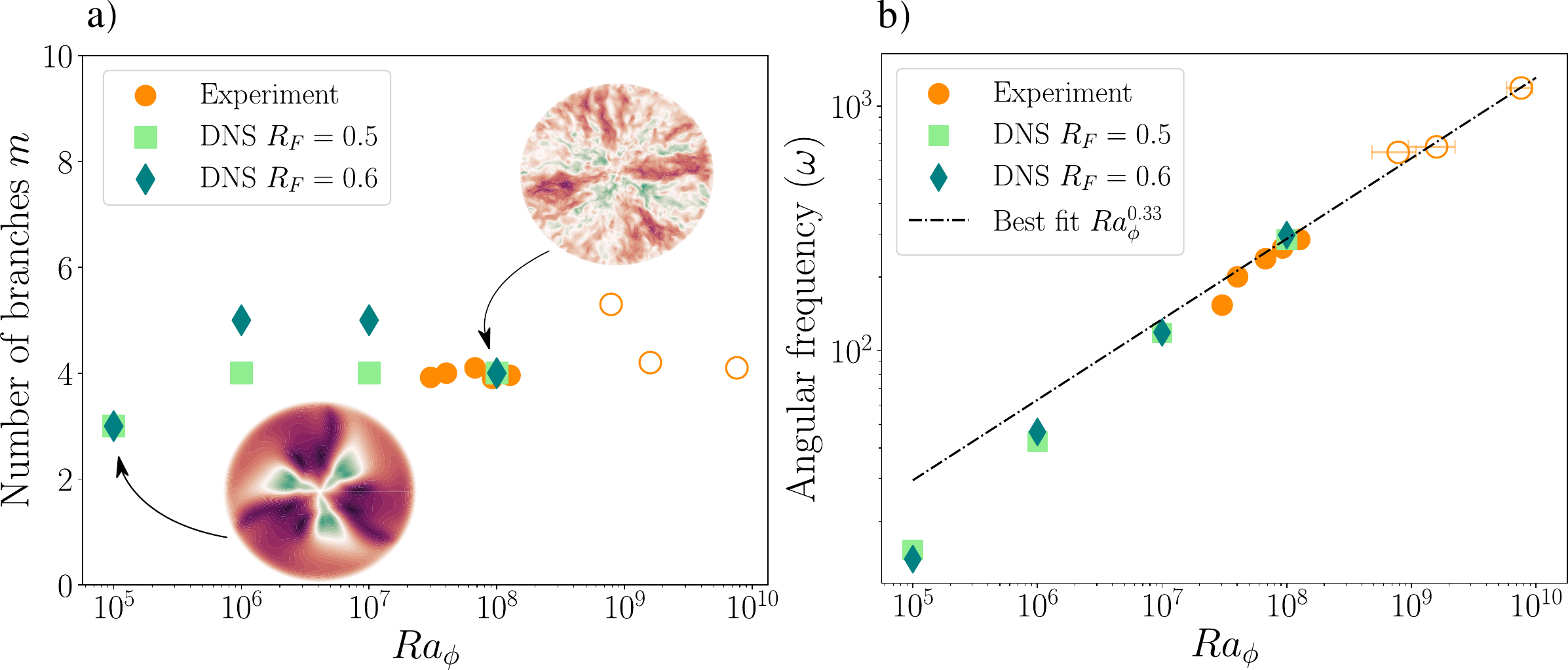}
   \caption{$a)$ Number of branches as a function of $\Ra$. The insets show two snapshots of the radial velocity maps from DNS at $z=0.66$ for $\Ra=10^5$ and $10^8$, $\RF=0.6$, $\Pra=0.7$, and $\Gamma=4$. $b)$ Angular drift frequency as a function of $\Ra$. The dashed-dotted line represents a best fit based on the experimental data for $Ra_\phi > 5 \times 10^7$. In both graphs, square symbols ($\square$), diamond symbols ($\Diamond$), and circle symbols ($\circ$) respectively represent data for DNS at $\RF=0.5$ and $\RF=0.6$, and experimental data. Empty circles indicate the use of \ch{SF6}. All DNS were performed with $\Pra=0.7$ and $\Gamma=4$. }
    \label{insta}
\end{figure}

The two determined frequencies are related by $\omega=m~\omega_p$, where $m$ is the number of branches or azimuthal wave number.
 Figure \ref{insta} shows $m$ and $\omega$ as functions of the Rayleigh number for all our experimental and numerical results. 

We note a good overall qualitative agreement, even if DNS exhibit a stronger dependence of $m$ on the flux ratio. 
Regarding $\omega$, a power law emerges, indicating a dependence  $\omega \sim \Ra^{1/3}$.
This relationship was determined through a best fit using the least squares method, applied to the experimental data while excluding values below $\Ra = 5\times10^{7}$  to focus on the asymptotic behavior at high Rayleigh numbers.
This $1/3$ exponent is reminiscent of the one identified for the horizontal velocity in Figure \ref{velocityandTresult}. However, we argue that both are not directly related. Indeed, from the DNS at $R_F=0.5$ and 0.6, we can compute the mean angular velocity averaged in $\varphi$ and $z$ at $r=3\Gamma/2$ (our conclusion remains qualitatively unchanged for other radii).
For the low $\Ra$ values of DNS, this advection angular velocity is significantly smaller than the measured angular drift velocity. Hence, 
we argue that the drifting pattern is not caused by the advection of some structure by a mean azimuthal velocity, but rather seems related to a wave-like phenomenon where the fluid, on average, does not undergo significant azimuthal movement.
We also note from our systematic DNS study that this pattern appears only above a given threshold in $\Ra$.
Initially, prograde and retrograde  similar structures superimpose.
Increasing $\Ra$, one of the two directions is randomly selected, and then persists up to the largest $\Ra$ investigated. 
A detailed study of the onset of this oscillatory instability is necessary.
This will be the subject of future work, including comparisons with other large-scale structures observed in turbulent convection, such as the Jump Rope Vortex observed in liquid metal by   \cite{Akashi_2022,Cheng_2022,Teimurazov_2023}.

\subsubsection{Analysis in the reference frame of the drifting pattern}

To gain deeper insights into the temperature and velocity field structure associated with the drifting pattern, we analyse our DNS in the frame of reference rotating with the pattern. To do so, we first determine the angular drift frequency ($\omega$) and azimuthal wavenumber ($m$) from the time and spatial spectra of the azimuthal velocity. Then, we spectrally interpolate, at each time step,  all variables on a grid ($X(t), Y(t), Z(t)$) rotating with the phase speed $\omega /m$ of the pattern.
We finally introduce a phase averaging operator denoted by $\langle \cdot \rangle_p$,  which is effectively a time average over a time $\tau_p$ within the reference frame of the drifting pattern, defined as follows:
\begin{equation}
\langle T \rangle_p = \frac{1}{\tau_p}\int^{t_0+\tau_p}_{t_0} T\left(t,X,Y,Z\right) \mathrm{d}t \ .
\end{equation}
The typical value of $\tau_p$ ranges between 0.2 and 1 diffusive times for the largest and the lowest Rayleigh numbers respectively. 
To ensure the convergence of the data obtained from this averaging, we verify that the probability density functions (PDF) of the radial temperature gradient and velocity fields remain consistent even when considering only half of the data. 

\begin{figure}
    \centering
    \includegraphics[scale=0.22]{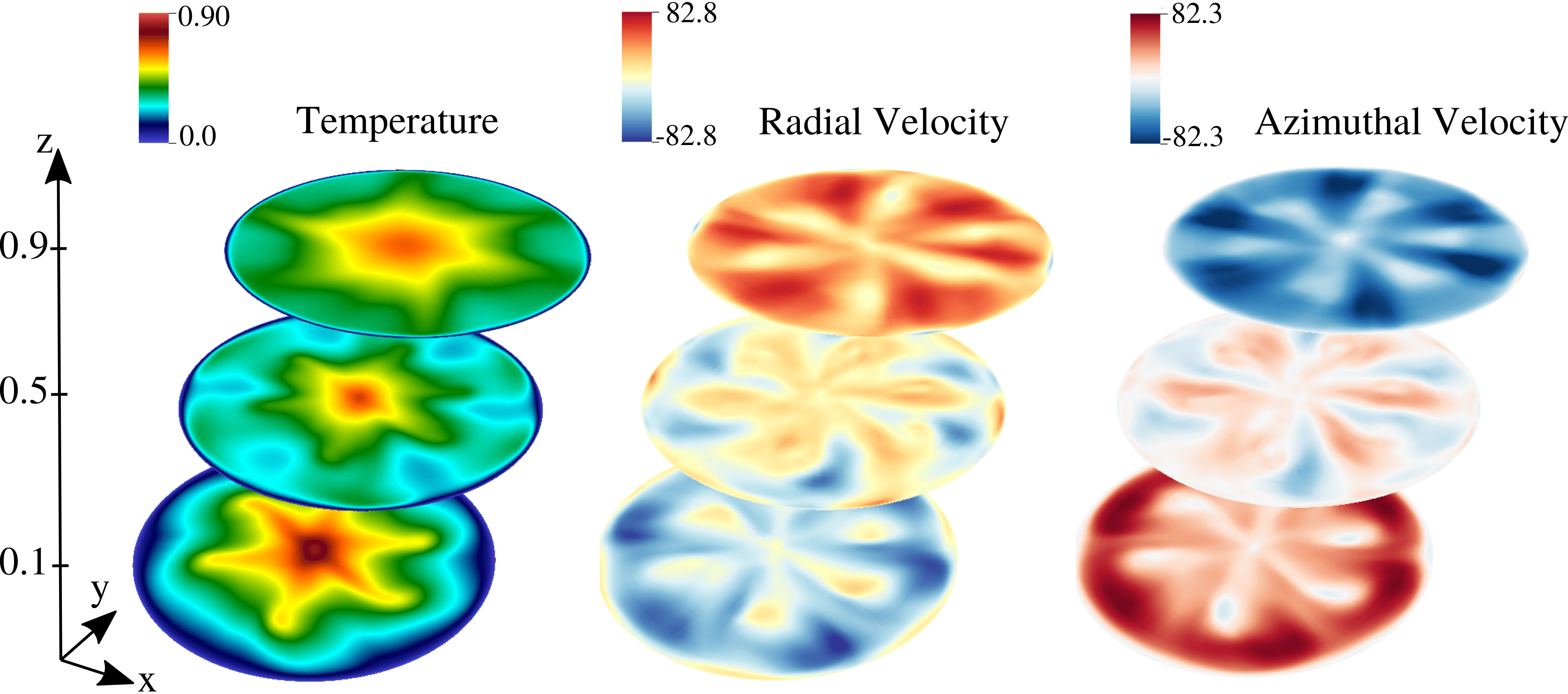}
   \caption{Phase-averaged temperature, radial velocity, and azimuthal velocity fields, processed during 1 diffusive time. Results are presented for three horizontal slices at $z=0.1$, $0.5$, and $0.9$. Parameters are $\Gamma=4$, $R_F=0.1$, $\Ra=10^5$ and $\Pra=0.1$. In the present DNS, the pattern drifts anticlockwise.}
    \label{PS}
\end{figure}

Figure \ref{PS} shows an example of the obtained  temperature, radial velocity and azimuthal velocity fields at three different depths. Thanks to the phase average, we now clearly identify the 6 branches pattern on all fields, while it would otherwise disappear when using a standard time average in the laboratory frame. 
Furthermore, in the upper part of the domain, we observe positive radial velocities moving from the center towards the edge, while in the lower part, the radial velocities are negative, moving from the edge towards the center. 
This aligns with the cooling effect applied at the edge, causing the fluid to cool, increase in density, and subsequently descend along the wall. 

Concerning the azimuthal velocity, a clear shear is evident between the upper ($z > 0.5$) and lower ($z < 0.5$) regions. 
This shear was unnoticed without the phase average, because it was masked by temporal fluctuations of similar magnitude to the average flow (not shown).
Note that we observe a shear in the opposite direction compared to Figure \ref{PS} when the pattern drifts clockwise.
In the pattern frame of reference, phase shifts between the fields are observed. The temperature and radial velocity seem to be in phase opposition, while the azimuthal velocity appears to be in phase quadrature with respect to both the temperature and radial velocity (not shown).

We see on the temperature field a well-defined 6 branches hot pattern at $z=0.1$, which becomes more diffuse as $z$ increases. Yet, for any given depth and radius, large azimuthal fluctuations are observed, including close to the outer wall. This thermal pattern thus plays a crucial role in the transport of the heat flux to the side, which will be the focus of the next section. 

\subsection{Wall flux fluctuations}

In line with Rayleigh-Bénard studies \cite[e.g.][]{Shang_2003,Shang_2004,chilla,Lakkaraju_2012,Labarre_2023}, we want to quantify heat flux fluctuations on the lateral wall and to understand their connection with both turbulent fluctuations and the large-scale flow structures.
To do so, we define the outgoing flux at the wall $r=\Gamma$ by 
\begin{equation}
   \Phi (t,\varphi,z)=-\frac{\partial T}{\partial r} \bigg|_{r=\Gamma}.
\end{equation}
Once the system reaches a statistically stationary state, the global flux balance 
(by integrating \eqref{nrj} over the volume) implies that the flux mismatch between the top and the bottom $(1-R_F)\pi\Gamma^2$ has to be balanced on time average by the outgoing flux at the side denoted by $\Phi_{\mathrm{side}}2\pi\Gamma$. Hence
\begin{equation}
    \Phi_{\mathrm{side}}=\frac{(1-R_F)\Gamma}{2} \ .
\end{equation}
For the following subsections, the wall heat flux $\Phi$ is normalized by this mean side flux $\Phi_{\mathrm{side}}$.

\subsubsection{DNS results} \label{sec:DNSres}
\begin{figure}
   \centering
    \includegraphics[scale=0.35]{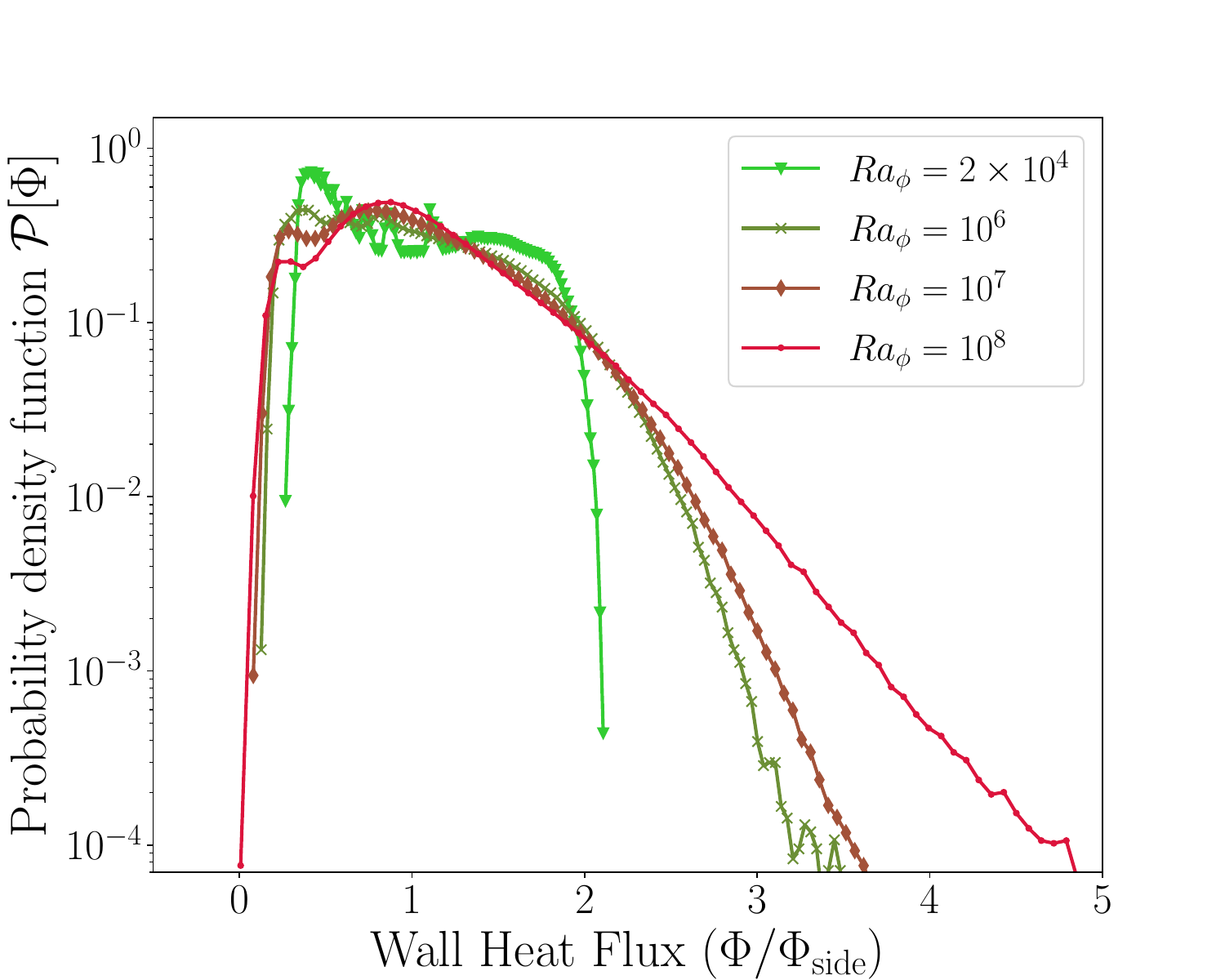}
   \caption{Wall heat flux PDF from DNS with  different values of $Ra_{\phi}$. The fluxes have been scaled by the time-averaged flux $\Phi_{\mathrm{side}}$. Input parameters are $\Pra=0.1$, $R_F=0.1$ and $\Gamma=4$.} 
    \label{figflux}
\end{figure}

In \autoref{figflux}, we analyze the flux fluctuations in DNS,  considering all azimuthal and vertical positions along the side wall over a period of $0.2$ thermal diffusive time. 
The PDFs of the heat flux are depicted for $Ra_{\phi}$ ranging from near the instability threshold at $\Ra=2\times10^4$ to the fully turbulent regime at $\Ra=10^8$. 
Note that the PDFs remain mostly unchanged when varying the aspect ratio from 4 to 16 at $\Ra=10^6$, $\Pra=0.1$ and $\RF=0.1$ (not shown).

With increasing $Ra_{\phi}$, more intense fluctuations across a wider range of values are observed. 
An exponential tail seems to emerge notably at $Ra_{\phi}=10^8$.
It is worth noting that \cite{Lakkaraju_2012} highlighted a similar exponential tail pattern in the heat flux along the sidewall of a cylindrical Rayleigh-Bénard system, with the slope increasing with $\Ra$. Additionally, \cite{Shang_2003,Shang_2004}, in a cartesian Rayleigh-Bénard configuration, identified a similar exponential behavior in the distribution of both vertical and horizontal bulk heat fluxes.
This exponential tail is not observed close to the onset of instability, but requires reaching sufficiently turbulent conditions.

The large-scale, drifting pattern is also responsible for intense fluctuations in heat flux near the wall.
\begin{figure}
    \centering
    \includegraphics[width=1\linewidth]{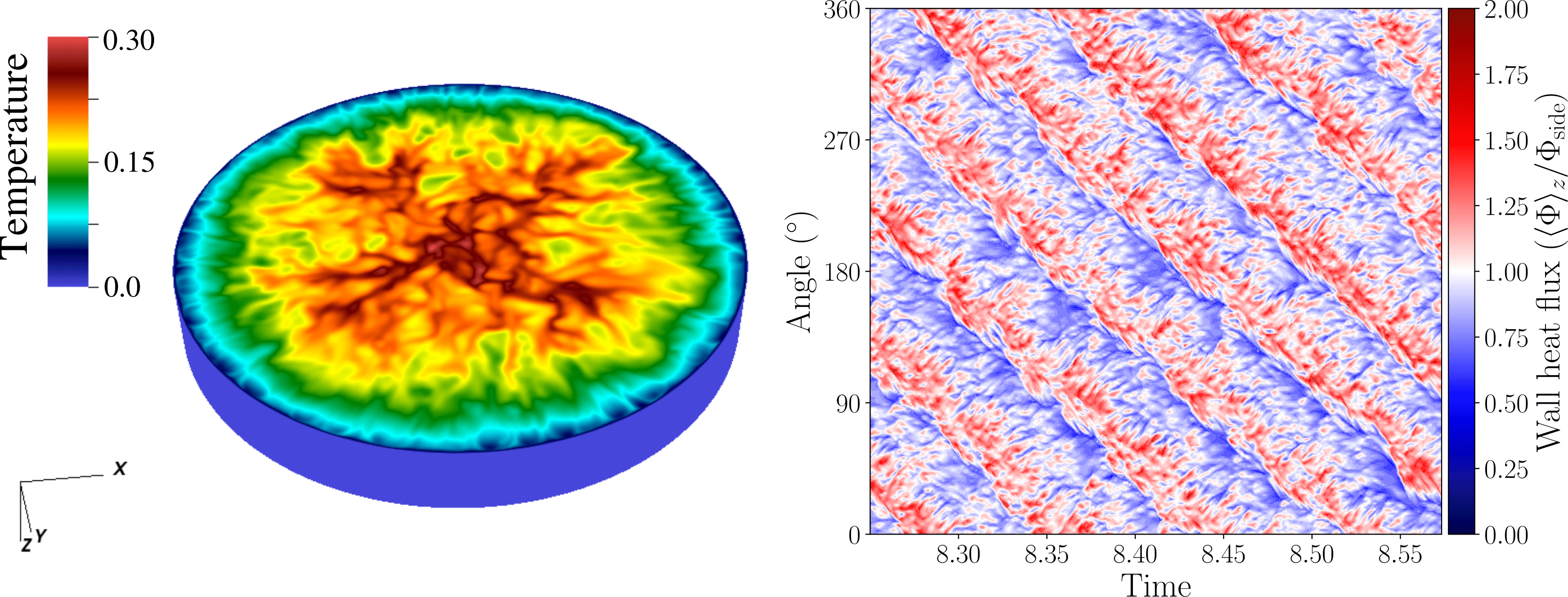}
   \caption{a) 3D view of the temperature field at the bottom plate (note the upside-down axes system with $z$ pointing downward) with $\Gamma = 4$, $\Ra=10^7$, $R_F=0.1$, and $Pr=0.1$. b) Hovmöller (space-time) diagram from this simulation of the wall heat flux averaged in $z$ and scaled by the mean wall flux $\Phi_{\mathrm{side}}$.}
   \label{mapflux}
\end{figure}

To illustrate this point, we performed a DNS at $\Ra=10^7$, $\Gamma=4$, with $\Pra$ = 0.1 and the flux ratio at $R_F=0.1$.
 \autoref{mapflux}a shows a snapshot of the temperature field, seen from below. The largest temperatures are localised along 4 drifting branches, as more clearly identified in the space-time diagram in \autoref{mapflux}b. 
Within these branches, intense fluctuations involve heat fluxes up to 2 times the average flux. 
Once scaled for our air experiment, the characteristic size and duration of one of these fluctuations are approximately \qty{4}{cm} and \qty{1.5}{s}, respectively.

\begin{figure}
   \centering
    \includegraphics[width=1\linewidth]{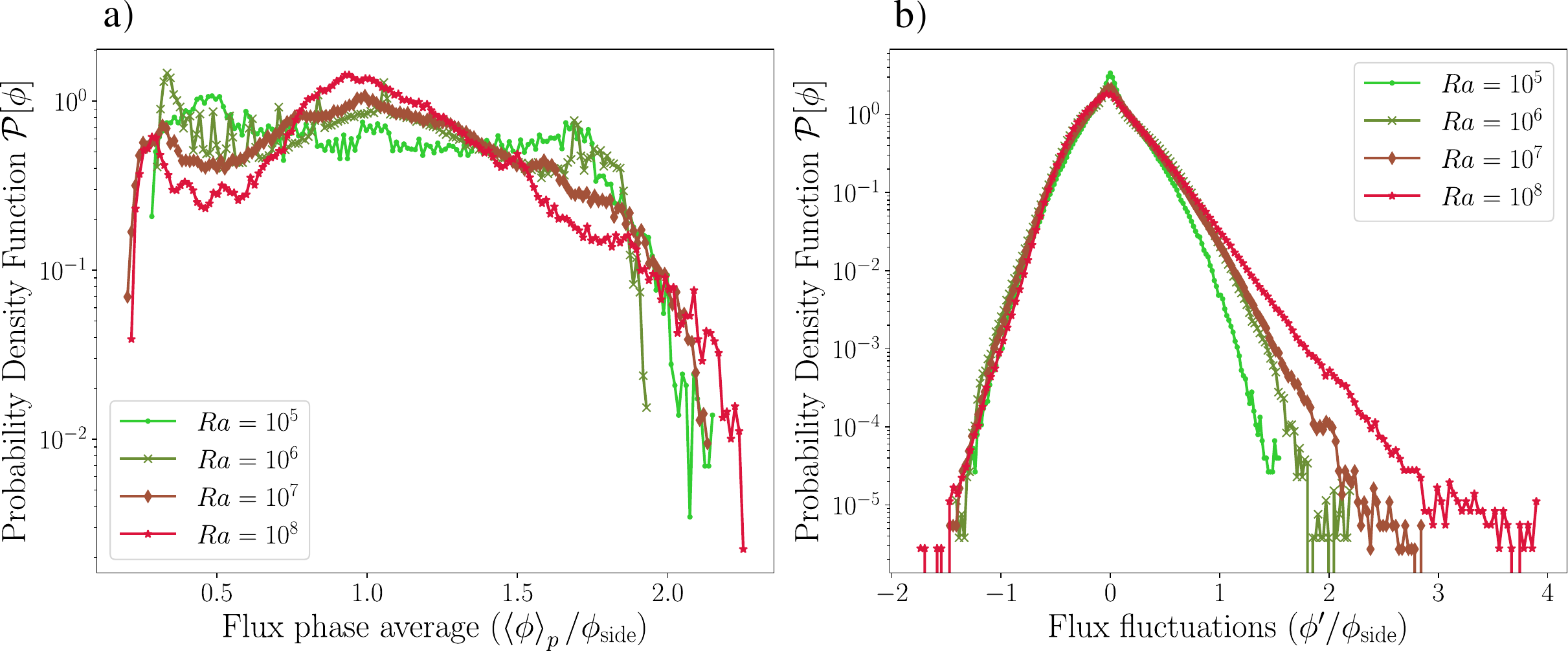}
   \caption{$a)$ PDF of the phase averaged wall heat flux as a function of $\Ra$. The other input parameters are $\Pra=0.1$, $R_F=0.1$ and $\Gamma=4$. All depths and azimuths along the side wall have been considered. $b)$ Same but for the wall heat flux fluctuations related to the phase average. For both plots, the fluxes have been scaled by the time-averaged flux $\Phi_{\mathrm{side}}$. } 
   \label{flux_phase_avg}
\end{figure}

We now want to disentangle the contributions to the side heat flux fluctuations from the large-scale drifting pattern and those from the small-scale turbulent convective fluctuations.
Figure \ref{flux_phase_avg}a shows the wall heat flux PDF after applying phase averaging: it thus focuses on the effect of the large-scale drifting pattern only. 
Note that similar analyses have been conducted in turbulent Rayleigh-Bénard systems in cylindrical geometry by \cite{Lakkaraju_2012} to assess how the local heat flux varies depending on the measurement location relative to the orientation plane of the Large Scale Circulation (LSC). 
We observe that the flux distribution carried by the pattern extends over a range from 0.5 to 2 times the mean flux $\Phi_{\mathrm{side}}$.
This range barely depends on the Rayleigh number.
Indeed, the drifting pattern is essentially responsible for the transport of the mean flux ($\Phi_{\mathrm{side}}$) out of the system, which is independent of $\Ra$. 
Increasing the $\Ra$ increases the rapid turbulent fluctuations which will be considered next.  
However, the shape of the PDFs changes with $\Ra$. Indeed, the observed  pattern evolves as $\Ra$ increases, notably by varying the number of branches and increasing the skewness of the flux profile associated with a branch (not shown). 

Figure \ref{flux_phase_avg}b) then shows the PDF of the wall heat flux fluctuations related to the phase averaged operator: we thus focus here on the turbulent fluctuations taking place around the phase average field.
As the Rayleigh number increases, the range of heat flux fluctuations widens.
We also see the emergence of the exponential tail at $\Ra=10^8$, previously identified in Figure \ref{figflux}.
This means that the largest and rarest heat flux events are associated with turbulent fluctuations, and not with the pattern itself.
Their probability will keep increasing with $\Ra$. 
It is also worth noting that the left portion of the fluctuations, with amplitude ranging from -2 to 0, appears to be unaffected by changes in the Rayleigh number, showing a consistent overlap. 
Since $\phi^\prime = \phi - \langle \phi \rangle_p$ and the heat flux is consistently positive for $R_F=0.1$, negative $\phi^\prime$ values suggest low heat fluxes $\phi$ and high heat fluxes caused by the pattern's drift $\langle \phi \rangle_p$. 
These occurrences indicate scenarios where the observed heat flux primarily arises from the pattern's drift.
This finding aligns with the observation that the pattern's drift consistently induces similar fluctuations regardless of the Rayleigh number (refer to Figure \ref{flux_phase_avg}a).

In summary, the heat flux fluctuations within the system exhibit a broad distribution that can be attributed to two main contributions: 
the pattern contribution, which transports the mean flux $\Phi_{\mathrm{side}}$ with a fixed amplitude independent of $\Ra$, and the fluctuations, which exhibit wider distributions and exponential tails as $\Ra$ increases.
In light of these numerical observations, how does experimental analysis contribute to our understanding of the heat flux fluctuations ?

\subsubsection{Experimental results}
 
\begin{figure}
    \centering
    \includegraphics[width=0.6\linewidth]{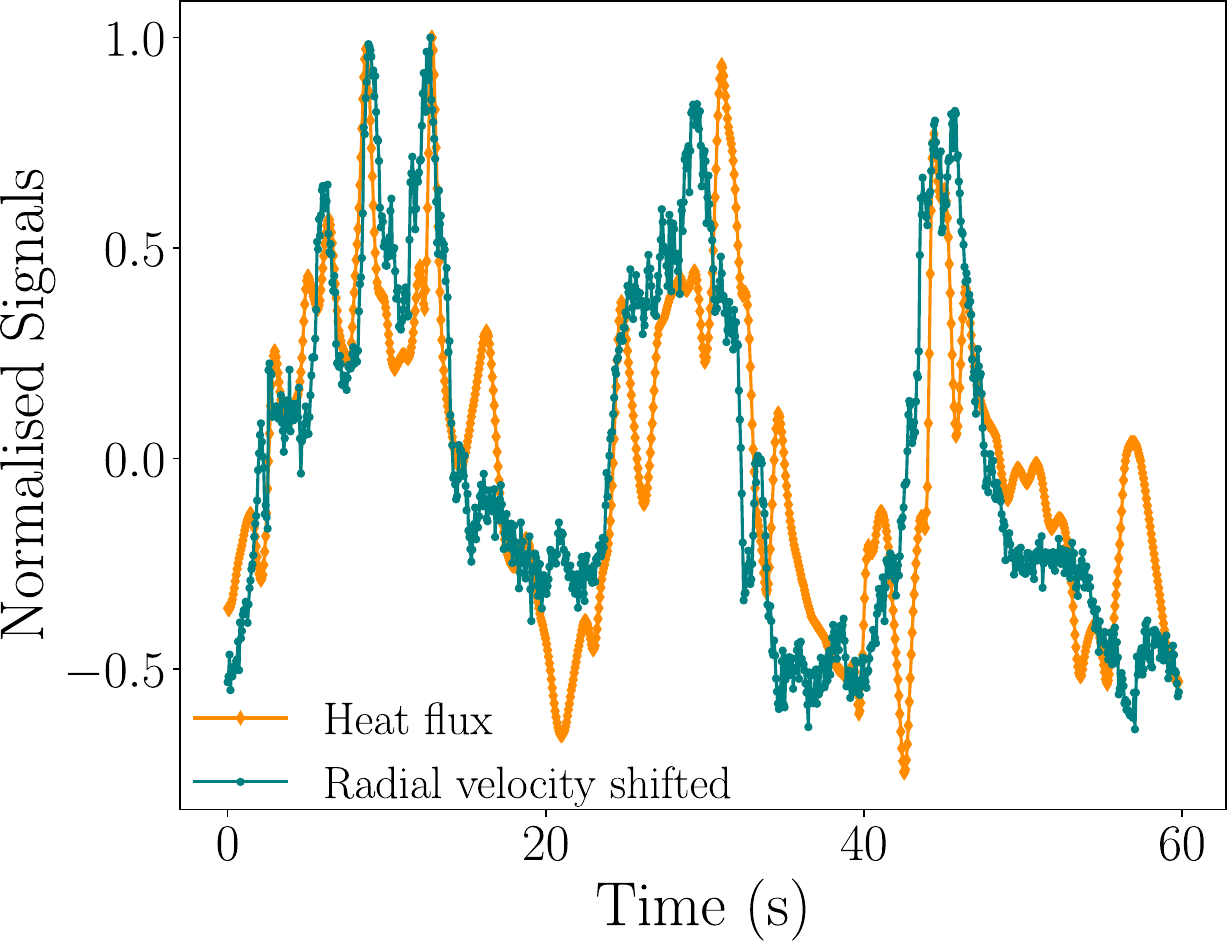}
   \caption{Temporal evolution of the wall heat flux measured by our probe spanning a width of $\qty{18}{mm}$ and a depth $z\in [\qty{4,6}{cm}, \qty{6,4}{cm} ]$ (orange), and radial velocity measured by PIV at $r=\qty{34}{cm}$,  $z=\qty{6}{cm}$, in the direction aligned with the flux sensor (green). The time is shifted by \qty{2.6}{s} for the velocity. For each signal, the mean temporal component has been removed to keep only the fluctuating part, which is then normalized by the maximum value of the signal. Data from experiment $n^o$= 3 in table \ref{table:1expe}.}
    \label{correlation}
\end{figure}
In \autoref{correlation}, we plot the time evolution of the wall heat flux measured experimentally as well as the radial velocity at a location situated \qty{10}{cm} from the flux sensor in the radial direction towards the center.
The purpose of this measurement is to explore the relationship between radial velocity and heat flux measured at the wall.
Upon examination, we observe that when we shift the radial velocity by a characteristic time $\tau_{\text{adv}}= \qty{2.6}{s}$, the two rescaled signals reasonably match. 
Moreover, the characteristic time $\tau_{\text{adv}}$ aligns with an advective timescale, namely $d/U$, where $d$ represents the distance between the flux sensor and the location of the velocity measurement, and $U$ corresponds to the time-average radial velocity at that specific location.
This alignment suggests a correlation between the radial velocity and the wall heat flux. 
This is compatible with the numerical results shown in Figure \ref{mapflux}b), and indicates that the drifting branches associated with higher radial velocity values act as pathways for transporting heat fluctuations towards the wall.

\begin{figure}
   \centering
    \includegraphics[scale=0.33]{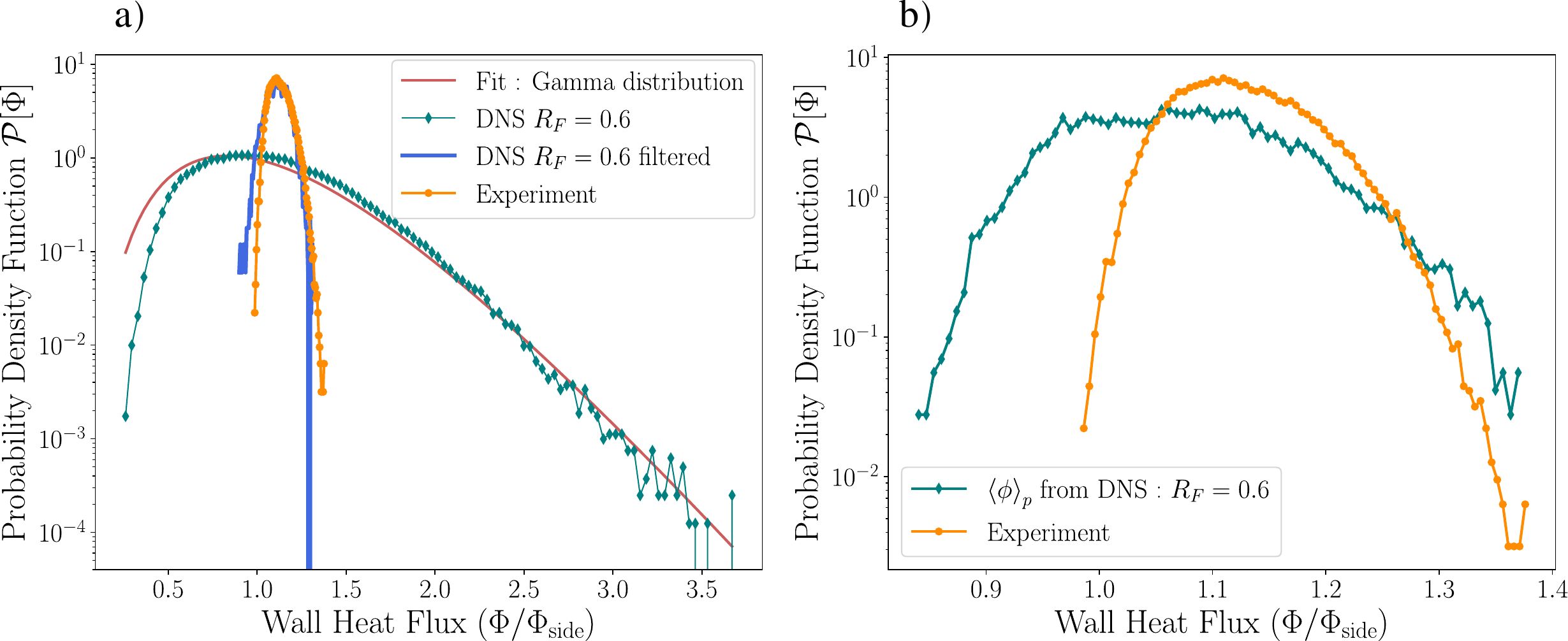}
   \caption{(a) Wall heat flux PDF for experiment $n^o 9$ in table \ref{table:1expe} (orange $\circ$) and for a DNS with $\Ra=10^8$, $\Gamma = 4$, $\Pra=0.7$, $R_F=0.6$ (green $\Diamond$). The continuous blue thick line represents the low-pass filtered DNS results. The thin red line represents a fit by a gamma distribution. The fluxes are all scaled by the time-averaged flux $\Phi_{\mathrm{side}}$. (b) Same for experiment $n^o 9$ and for DNS  heat flux fluctuations due to the drifting pattern only (see section \ref{sec:DNSres}).  } 
   \label{pdf_manip}
\end{figure}

In \autoref{pdf_manip}a), we compare the PDF of the wall heat flux derived from experimental measurements and from numerical simulations.
The heat flux is measured experimentally at mid-height and averaged over the entire surface of the sensor. It is positioned halfway up the sidewall, extending end-to-end from $z=\qty{4.6}{cm}$ to $z=\qty{6.4}{cm}$, with a width of $\qty{1.8}{cm}$.
We consider experiment $n^o 9$ from table \ref{table:1expe}, where the non-dimensional parameters have been estimated by $\Ra=1.26\times 10^8$ and $R_F=0.54$.
Numerically, we fix $\Ra=10^8$, $\Gamma=4$, $\Pr=0.7$ and consider the flux ratio $R_F=0.6$.
Given the numerical difficulty to get statistics over long periods, we assume system ergodicity.

As a result, the numerical flux calculations consider all azimuthal events at each output snapshot, limited in height to match the depth of the flux sensor ($z\in[\qty{4.6}{cm}$ to $ \qty{6.4}{cm})]$). 

When comparing the PDFs, we initially note that the numerical PDF predicts much larger fluctuations than those measured experimentally. 
However, it should be noted that the sensor has a response time ($T_r$) of 0.7 second, meaning that any event occurring over a period shorter than this will not be directly measured.
To account for this experimental aspect, we apply a first-order low-pass temporal filter, with a cutoff frequency ($f_c$) corresponding to the flux sensor's response time ($f_c=1/T_r$). 
The application of this filter is done in Fourier space by multiplying the Fourier transform of the experimental signal by the transfer function of a first-order low-pass filter.
We then note a good qualitative concordance between DNS and experimental results. 
A priori, a 2D spatial filter should be considered, since the heat sensor averages over its entire surface.
However, the good agreement suggests this is unnecessary, as short events correlate with small-scale phenomena.
Notice that we did the same numerical computation but for
$R_F=0.5$, and the same qualitative agreement was found.
As the result, the most extreme events then appear to have a relatively brief duration, at least shorter than the response time associated with the sensor.

The shape of the numerical PDF of \autoref{pdf_manip}a) is compatible with a gamma distribution, denoted $\mathcal{P}_{\gamma}[\Phi]$ and defined by :   
\begin{equation}
    \mathcal{P}_{\gamma}[\Phi] = \frac{1}{\gamma(N) \Psi^N}~\Phi^{N-1} e^{-\Phi/\Psi}.
\end{equation}
Here, $\gamma(N)$ represents the gamma function with $N$ and $\Psi$ the 2 fitting parameters. 
$N$ is an integer without dimensionality, while $\Psi$ has a dimension of a heat flux. Since in our case the mean flux is imposed, $\Psi$ is constrained to $\Phi_{\mathrm{side}}/N$. 
The distribution now has only one free parameter, $N$, which is  set to 6 in \autoref{pdf_manip}a), but depends on the input parameters $\Ra$, $R_F$, $Pr$, and $z$ due to vertical inhomogeneity.
In \autoref{figflux} for example, the PDF accounting for all events along the vertical direction $z$ at $R_F=0.1$ and  $\Ra=10^8$ is best fitted by a gamma distribution with $N=4$.

In \autoref{pdf_manip}b), we analyze the PDF of wall heat flux, comparing experiment $n^o  9$ with the same DNS conducted at $Pr=0.7$, $\Gamma=4$, $R_F=0.6$, and $\Ra=10^8$, but considering only the heat flux distribution due to the drifting pattern. 
Both PDFs are similar. This makes sense since the drifting pattern is persistent and drifts slowly: it is hence easily catched by the sensor despite its limited response time. The heat sensor however misses the rapid, turbulent fluctuations described in section \ref{sec:DNSres}. 
The experimental PDF also appears to have fewer values with  $\Phi/\Phi_{\mathrm{side}} < 1$ compared to the DNS results.
While a perfect match between the two PDFs is not expected, the remaining discrepancy likely stems from the fact that the flux sensor averages over its surface, a factor not accounted for in the numerical data.

\section{Conclusion and future work}

In conclusion, the experiments presented in the present paper allowed to observe a $\Ra^{-1/5}$ scaling law for the mean temperature, consistent with the scaling identified in \cite{reinjfm,reinnureth}, and extended over nearly 2 orders of magnitude in Rayleigh number, reaching $\Ra \simeq 10^{10}$  values commensurate with those of the nuclear safety application  that initially motivated our work.
Additionally, a $\Ra^{1/3}$ scaling law for the RMS horizontal velocity has been identified.
This behavior is explained through dimensional analysis, suggesting that the scaling originates from horizontal heat transport in the turbulent regime.
In addition to scaling laws for globally averaged quantities including the RMS velocity and the temperature, experimental measurements have confirmed, even for the largest values of $\Ra$, the presence of a mesoscale, persistent branch pattern associated with large-scale, slow wall heat flux fluctuations.
Those superimpose to small-scale, rapid, high amplitude fluctuations associated to turbulent fluctuations. 

Regarding the nuclear safety application, we are interested in understanding heat transfer through the studied metal layer towards the lateral metal wall, in order to predict its melting leading possibly to vessel failure. We argue that the turbulent fluctuations may be secondary: indeed, even if they involve events associated with the most extreme fluxes, they manifest in very localized spatial areas and for short durations, hence correspond energetically to weak events. 
On the other hand, the thermal branches, associated with slower and larger events, play a significant role in heat exchanges, as they cover substantial portions of the wall and persist for longer durations.

To further explore the influence of these fluctuations on the structural integrity of the wall, it would be valuable to examine a global melting scenario, incorporating both convective flow and melting phenomena.
This sort of challenging problem has been investigated across different domains, spanning geophysics \citep{Alboussière2010,Labrosse_2018} and industrial applications \citep{Kalhori_1985,fragnito_2022}. Published studies encompass Rayleigh-Bénard convection \citep{Favier_2019,Labrosse_2018} as well as natural convection \citep{Jany_1988,Viskanta_1988,Lacroix_1993}, particularly relevant to our case.
The characteristic feature of our problem is the periodic heat forcing due  to the pattern rotation, which may lead to various melting regimes  
\citep{Ho_chu_1993}.

From a fundamental point of view, the next step is to tackle the instability at the origin of the drifting pattern. 
Physical insights into the underlying  mechanism could indeed be useful to better forecast the heat flux fluctuations. This will be the focus of our future work.

\section{Declaration of Interests}
The authors declare the following interests regarding the funding and support received for this work: L'institut de radioprotection et de sureté nucléaire (IRSN), Commissariat à l'énergie atomique et aux énergies alternatives (CEA) and  Electricité de France (EDF) provided financial support for this project. The aforementioned organizations had no influence on the design, data collection, analysis, interpretation of results, or the decision to publish. The content of this work remains the sole responsibility of the authors.

\newpage
\section{Appendix}

\subsection{Summary of the experiments/simulations parameters} \label{annexe}
Tables \ref{table:1expe} and \ref{table:1num} provide all relevant parameters for the performed experiments and DNS, respectively.

\begin{table}
\centering
\begin{tabular}{ccccccccc}
\noalign{\global\arrayrulewidth=0.01mm}
\multicolumn{1}{c}{$n^o$} & \multicolumn{1}{c}{$P(\mathrm{W})$} & \multicolumn{1}{l}{Gas} & \multicolumn{1}{c}{$\rho_{\mathrm{eff}}  (\qty{}{kg/m^3})$} &  \multicolumn{1}{l}{$\langle \theta \rangle$ $(\qty{}{\degreeCelsius})$} & \multicolumn{1}{c}{$R_F$} & \multicolumn{1}{c}{$Ra_{\phi}$} & \multicolumn{1}{c}{$Pr$} & \multicolumn{1}{c}{$\Gamma$} \\ \hline
$1$ & $75$ & Air & 1.2 & 28.7 &$0.43$ & $3.2\times10^7$ & 0.71 & 4 \\
$2$ & $100$ & Air & 1.2 & 30.5 &$0.44$ & $4.0\times10^7$ & 0.71 & 4 \\
$3$ & $100$ & Air & 1.2 & 31.4 &$0.48$ & $4.6\times10^7$ & 0.71 & 4 \\
$4$ & $150$ & Air & 1.2 & 35.6 &$0.57$ & $6.8\times10^7$ & 0.71 & 4 \\
$5$ & $150$ & Air & 1.2 & 35.3 &$0.58$ & $7.0\times 10^7$ & 0.71 & 4 \\
$6$ & $150$ & Air & 1.2 & 36.6 &$0.54$ & $6.75\times 10^7$ & 0.71 & 4 \\
$7$ & $200$ & Air & 1.2 & 40.4 &$0.57$ & $9.26\times 10^7$ & 0.71 & 4 \\
$8$ & $200$ & Air & 1.2 & 42.3 &$0.54$ & $9.88\times 10^7$ & 0.71 & 4 \\
$9$ & $250$ & Air & 1.2 & 48.2 &$0.54$ & $1.26\times 10^8$ & 0.71 & 4 \\
$10$ & $100$ & Air + \ch{SF6} & $2.9$ & 32.1 & $0.54$ & $8.02\times 10^8$ & $0.87$ & 4 \\
$11$ & $170$ & Air + \ch{SF6} & $3.4$ & 38.6 & $0.57$ & $1.61\times 10^9$ & $0.87$ & 4  \\
$12$ & $400$ & Air + \ch{SF6} & $4.1$ & 61.5 & $0.56$ & $7.6\times 10^9$ & $0.85$ & 4 \\
\end{tabular}
\caption{Experiment parameters for the 12 performed cases. P is the injected power in watts (W).}
\label{table:1expe}
\end{table}
\begin{table}
\centering
\begin{tabular}{cccccccccc}
\noalign{\global\arrayrulewidth=0.01mm}
\multicolumn{1}{c}{$Ra_{\phi}$} & \multicolumn{1}{c}{$\Gamma$} & \multicolumn{1}{l}{$Pr$} & \multicolumn{1}{c}{$R_F$} & \multicolumn{1}{c}{DNS/filtered} & \multicolumn{1}{c}{$\mathcal{E}$} & \multicolumn{1}{c}{$N$} & \multicolumn{1}{c}{$\eta_K /L$}  & \multicolumn{1}{c}{BC}\\ \hline
$10^2$ & $[4;8;16]$ & $0.1$ & $[0.1;0.9]$ & DNS & $[2688;9216;33792]$ & $8$ & $[5.8;4.4;3.1]$& FS\\
$10^3$ & $[4;8;16]$ & $0.1$ & $[0.1;0.9]$ & DNS & $[2688;9216;33792]$ & $8$ & $[5.8;4.4;3.1]$& FS\\
$10^3$ & $[4;8]$ & $0.1$ & $0.5$ & DNS & $[2688;9216]$ & $8$ & $[6.2;3.7]$& FS\\
$10^4$ & $[4;8;16]$ & $0.1$ & $[0.1;0.9]$ & DNS & $[2688;9216;33792]$ & $10$ & $[3.3;2.7;2.3]$& FS\\
$10^4$ & $[4;8]$ & $0.1$ & $0.5$ & DNS & $[2688;9216]$ & $10$ & $[3.3;2.5]$ & FS\\
$3.10^4$ & $[4;8;16]$ & $0.1$ & $0.1$ & DNS & $[2688;9216;33792]$ & $10$ & $[2.5;2;1.8]$& FS\\
$10^5$ & $[4;8;16]$ & $0.1$ & $[0.1;0.9]$ & DNS & $[2688;9216;33792]$ & $10$ & $[1.88;1.56;1.52]$& FS\\
$10^5$ & $4$ & $0.7$ & $[0.5;0.6]$ & DNS & $9216$ & $10$ & $4$ & NS\\
$10^5$ & $[4;8]$ & $0.1$ & $0.5$ & DNS & $[2688;9216]$ & $10$ & $[1.87;1.52]$ & FS\\
$3.10^5$ & $[4;8;16]$ & $0.1$ & $0.1$ & DNS & $[2688;9216;33792]$ & $10$ & $[1.43;1.35;1.36]$& FS\\
$10^6$ & $[4;8;16]$ & $0.1$ & $[0.1;0.9]$ & DNS & $[9984;33608;33792]$ & $10$ & $[4;3.3;1.22]$& FS\\
$10^6$ & $4$ & $0.7$ & $[0.5;0.6]$ & DNS & $9216$ & $10$ & $2.2$ & NS\\
$10^6$ & $[4;8]$ & $0.1$ & $0.5$ & DNS & $[9216;33608]$ & $10$ & $[3.7;3.22]$ & FS\\
$3.10^6$ & $[4;8;16]$ & $0.1$ & $0.1$ & DNS & $[9984;33608;33792]$ & $10$ &  $[3;2.5;1.12]$& FS\\
$10^7$ & $[4;8;16]$ & $0.1$ & $[0.1;0.9]$ & DNS & $[9984;33608;36608]$ & $10$ & $[2.2;1.8;1]$& FS\\
$10^7$ & $4$ & $0.7$ & $[0.5;0.6]$ & DNS & $9984$ & $10$ & $1.3$ & NS\\
$10^7$ & $[4;8]$ & $0.1$ & $0.5$ & DNS & $[9984;33608]$ & $10$ & $[2.3;1.74]$ & FS\\
$3.10^7$ & $4$ & $0.1$ & $0.1$ & DNS & $9984$ & $10$ & $1.8$& FS\\
$3.10^7$ & $[8;16]$ & $0.1$ & $[0.1;0.9]$ & filtered & $[33792;36608]$ & $10$ &  $[1.44;0.78]$ & FS\\
$10^8$ & $4$ & $0.1$ & $[0.1;0.9]$ & DNS & $9984$ & $10$ & $1.27$& FS\\
$10^8$ & $[4;8]$ & $0.1$ & $0.5$ & [DNS;filtered] & $[9984;33792]$ & $10$ & $[1.4;1.3]$ & FS\\
$10^8$ & $4$ & $0.7$ & $[0.5;0.6]$ & DNS & $33608$ & $10$ &  $2.5$ & NS\\
$10^8$ & $[8;16]$ & $0.1$ & $[0.1;0.9]$ & filtered & $[33792;36608]$ & $10$ & $$[1.33;0.62]$$ & FS\\
$3.10^8$ & $[4;8]$ & $0.1$ & $[0.1]$ & filtered & $[9984;33792]$ & $10$ & $[1.1;1]$ & FS\\

\end{tabular}
\caption{Simulations summary (DNS or filtered) according to the physical and numerical parameters. $\mathcal{E}$, $N$, BC and $\eta_K /L$ represent respectively the number of hexahedral elements, the polynomial spectral order, the top boundary conditions applied (FS--Free-slip and NS--No-slip) and the ratio between the Kolmogorov dissipative scale ($\eta_K$) and the numerical grid size ($L$).
For further details, readers are directed to \cite{reinjfm}.}
\label{table:1num}
\end{table}
\subsection{Estimation of the non-dimensional numbers within a gas mixture }\label{gasmixture}
For reaching the most extreme values of $\Ra$ experimentally, we use \ch{SF6}. However, \ch{SF6} is not pure in the system; it is mixed with air. 
The evaluation of the Rayleigh and Prandtl numbers therefore requires a precise knowledge of the physical properties of the air/\ch{SF6} mixture under consideration. 
For a perfect gas mixture, the mass density ($\rho_m$) as well as the specific heat capacity ($c_{p_m}$) of the mixture are linearly expressed as a function of the mass fraction of \ch{SF6} (denoted $y$), as follows: 
\begin{equation}
\left\{
    \begin{array}{ll}
       \rho_m =& (1-y) \rho_{\text{air}}+y\rho_{\ch{SF6}} \\ 
        c_{p_m} =& (1-y) c_{p_{\text{air}}}+yc_{p_{\ch{SF6}}}.
    \end{array}
\right.
\label{lin}
\end{equation}
Since the densities of \ch{SF6} and air are tabulated as functions of temperature and pressure conditions, the experimental measurement of the mean density $\rho_m$ enables the determination of the \ch{SF6} mass fraction. 
This, in turn, allows for the estimation of the heat capacity of the gas mixture.
Regarding dynamic viscosity and thermal conductivity, kinetic theory of gases allows for the determination of these properties \citep{chapman}, but it is difficult to apply because it involves numerous degrees of freedom. 
Simpler models allow for a relatively accurate estimation of these quantities. 
The model proposed by \cite{wilke} provides an estimation of the viscosity of a gaseous mixture (tested up to 7 species) accurate to \qty{1.9}{\%}. 
\begin{figure}
    \centering
    \includegraphics[scale=0.3]{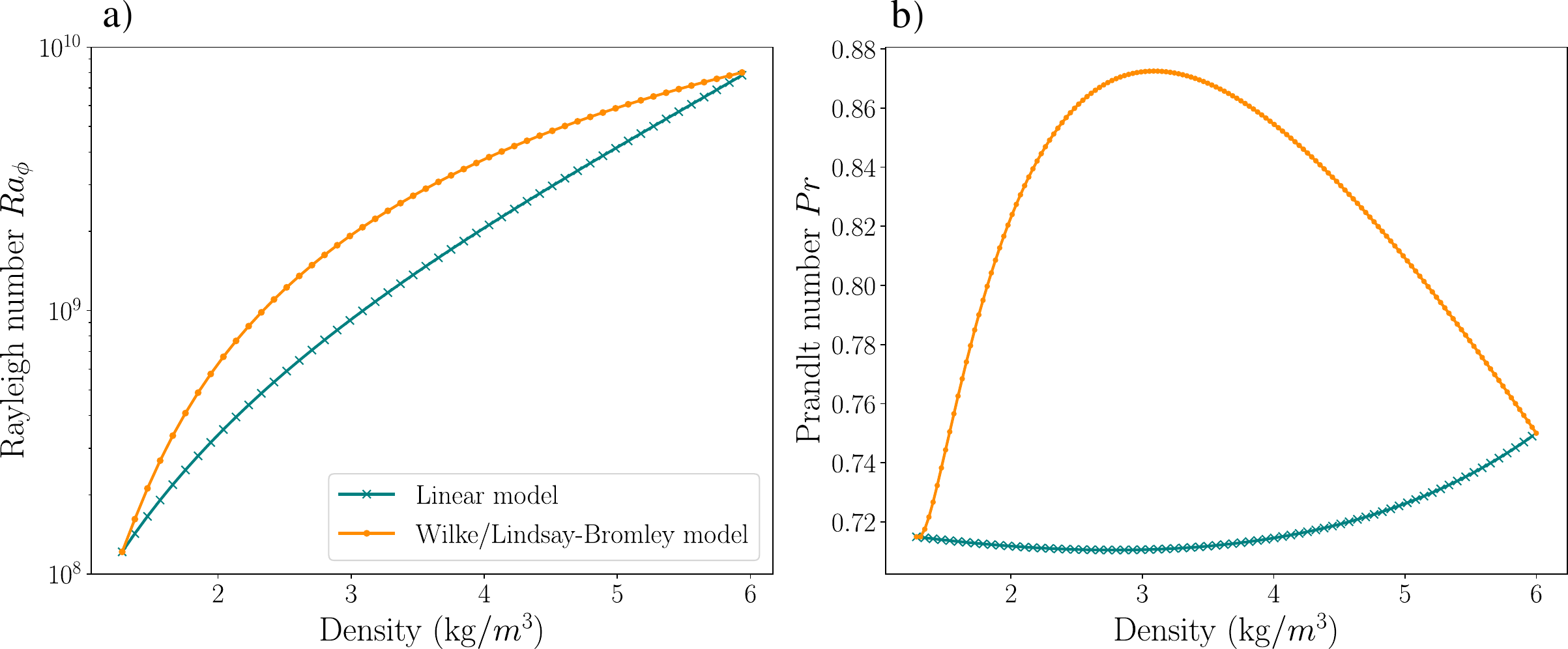}
   \caption{Evolution of $a)$ $\Ra$  and $b)$ $Pr$ as a function of the density of the air/\ch{SF6} mixture, considering either a linear mixture model for the viscosity and thermal conductivity or the \cite{wilke}/ \cite{lindsay} model.
            In all cases, $\phi=\qty{200}{W/m^2}$, and $H=\qty{11}{cm}$.}
    \label{mixture}
\end{figure}

The viscosity of the air/\ch{SF6} mixture, denoted as $\mu_m$, is expressed as follows:
\begin{equation}
    \mu_m= \frac{\mu_{\text{air}}}{1+A\frac{x_{\ch{SF6}}}{x_{\text{air}}}}  + \frac{\mu_{\ch{SF6}}}{1+\frac{1}{A}\frac{x_{\text{air}}}{x_{\ch{SF6}}}}
\end{equation}
where $x_{\text{air}}$ and $x_{\ch{SF6}}$ represent the mole fraction of air and of \ch{SF6}. 
$A$ is the Sutherland constant \citep{sutherland},  $A=\left(\frac{M_{\ch{SF6}}}{M_{\text{air}}} \frac{\mu_{\text{air}}}{\mu_{\ch{SF6}}} \right)^{1/2}$, where $M_{\ch{SF6}}$ and $M_{\text{air}}$ are the molar masses of \ch{SF6} and air. 
Concerning thermal conductivity, a similar model has been proposed by  \cite{lindsay}.
The difference between the Wilke model (viscosity) and the Lindsay and Bromley model (conductivity) lies in a variation of the Sutherland constant.

Figures \ref{mixture}a) and b) show how the Rayleigh number and the Prandtl number evolve as a function of the system's density. 
This evolution is studied using two different models: one with linear models for conductivity and viscosity (similar to \eqref{lin}), and the other where the Wilke model (viscosity) and Lindsay/Bromley model (conductivity) are used to estimate these properties. 

Significant deviation between these two estimates is observed for
densities around $\qty{3}{kg/m^3}$, with a deviation exceeding \qty{50}{\%} for the Rayleigh number and approximately \qty{20}{\%} for the Prandtl number.
This large Rayleigh number deviation is mainly due to its quadratic dependency on conductivity. 
Henceforth, the Wilke and Lindsay/Bromley models are used for estimating the viscosity/conductivity of the air-\ch{SF6} mixture.

\subsection{Experimental setup : energy balance}\label{energybalance}
In this section, we evaluate the thermal balance of the experimental setup in order to quantify both the thermal losses and the time-averaged ratio between the imposed heat flux and the one escaping through the lid ($\RF$). This last estimate is necessary for comparison with numerical simulations.
\subsubsection{Input heat flux and losses}
\label{fluxentrant}
\begin{figure}
    \centering
    \includegraphics[width=0.55\linewidth]{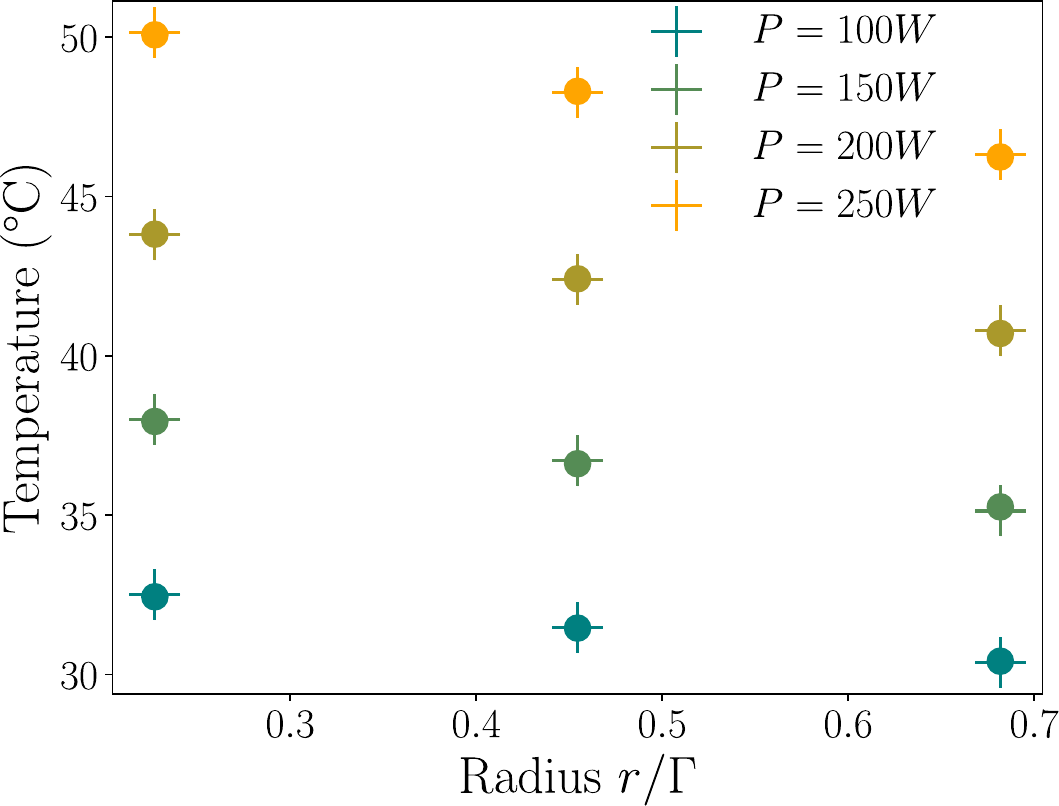}
   \caption{Average temperature as a function of radius for angular positions $\varphi=\ang{0}$ ($\bullet$) and $\varphi=\ang{90}$ ($+$), at different powers ranging from \qty{100}{W} to \qty{250}{W}.}
    \label{checkflux}
\end{figure}
In Figure \ref{checkflux}a, the temperature field from the camera, averaged over these 3 minutes, shows perfect uniformity, indicating a uniform heat flux.
First, to ensure system azimuthal uniformity, we use thermal measurements from the thermocouples. In Figure \ref{checkflux}b, the time-averaged temperature of each thermocouple is plotted against its location radius, for two specific angular positions $\varphi = \ang{0}$ ($\bullet$) and $\varphi= \ang{90}$ ($+$).
These measurements were taken for different powers, ranging from \qty{100}{W} to \qty{250}{W}. The average temperatures are independent of their angular position, confirming the azimuthal homogeneity of the system.

Now, let us evaluate the heat losses in our setup. When applying electrical power $P$ to the 8 heating elements, most of the flux is supplied to the aluminum plate, but a fraction escapes into the insulating plate. 
Let us introduce the heat efficiency coefficient, denoted $\chi$ and defined by $\chi=\phi_{\text{in}}/\phi_{\text{set}}$, where $\phi_{\text{in}}$ and $\phi_{\text{set}}$ represent the heat flux supplied to the aluminum plate and to the heat mats (calculated as $P/8S_{\text{heat}}$, with $S_{\text{heat}}$ as the heat mat surface area).
In the following, we outline 2 methods to estimate $\chi$.

The first method, named the \textit{transient method}, involves deriving the equation governing the time evolution of the aluminum plate temperature from an energy balance on the plate. 
The change in internal energy of the plate is due to the heat flux imposed by the heating elements, as well as the convective flux of the gas at the plate surface.
This is written as 
\begin{equation}
\rho_{\text{Al}} c_{p_{\text{Al}}} e_{\text{Al}} \frac{\mathrm{d}\theta}{\mathrm{d}t}=\phi_{\text{in}} - h\left(\theta-\theta_{\text{gas}}\right), 
\label{bilanalu}
\end{equation}
where $\rho_{\text{Al}}$, $c_{p_{\text{Al}}}$, $e_{\text{Al}}$, $\theta$ respectively represent the density, specific heat capacity, thickness, and temperature (assumed uniform) of the aluminum plate. 

\begin{figure}
    \centering
    \includegraphics[width=1\linewidth]{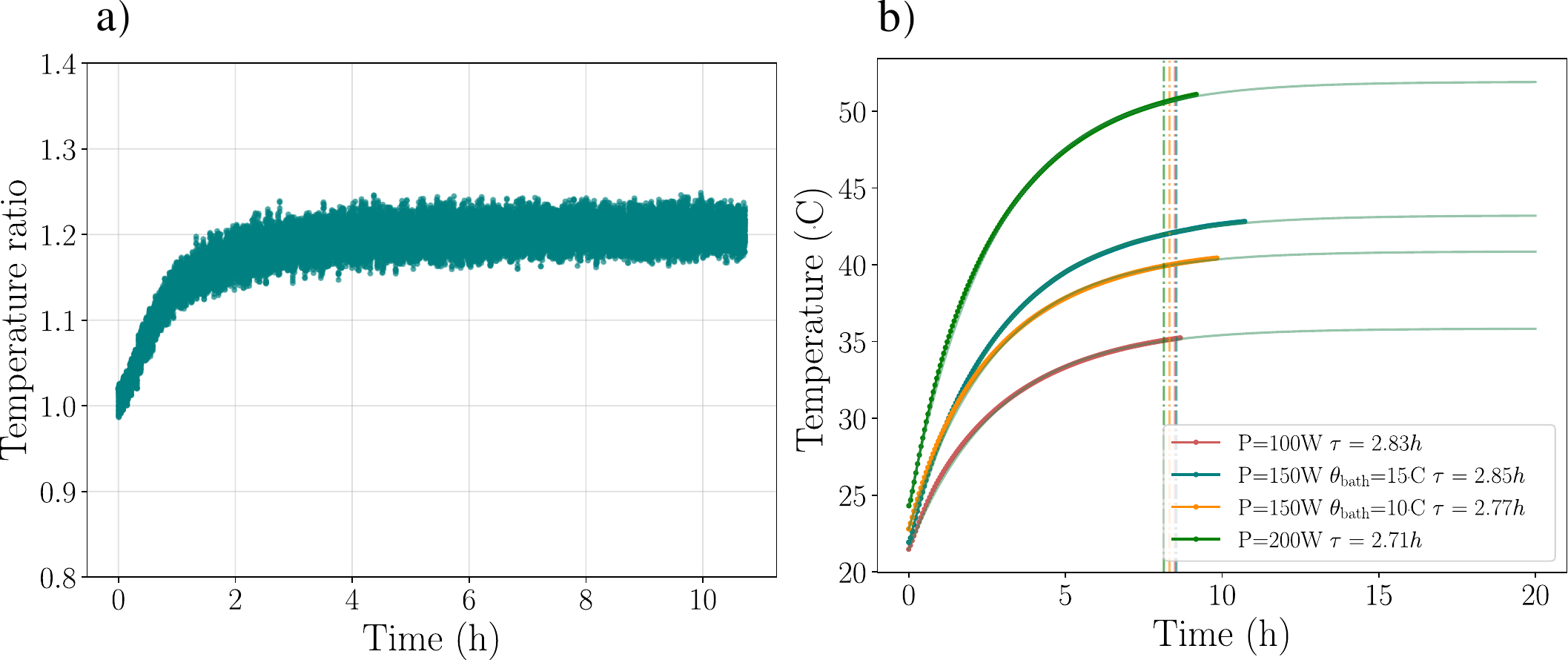}
   \caption{$a)$ Time evolution of the ratio between the temperature of the aluminum plate and the average temperature within the system, averaged over the six thermocouples ($\theta/\theta_{\text{gas}}$). The set power is $P =\qty{150}{W}$ and the bath setpoint temperature is $\theta_{\mathrm{bath}} =\qty{15}{\degreeCelsius}$. $b)$ Temporal evolution of the temperature at the center of the aluminum plate measured via the PT 100 positioned just tangent to the plate, for various powers and bath temperatures ($\theta_{\mathrm{bath}}$). Vertical lines indicate when time reaches 3 $\tau$, and solid green lines represent the best fit of the data using \eqref{solTalu} as a model. }
    \label{PT100plaque}
\end{figure}

Additionally, $h$ and $\theta_{\text{gas}}$ represent the convective heat transfer coefficient between the gas and the plate, and the ambient gas temperature.
It should be noted that $\theta_{\text{gas}}$ is not constant and increases during the heating period.
However, after a short transient phase, \autoref{PT100plaque}a) demonstrates that the ratio between the temperature of the aluminum plate measured at the center and the average temperature within the system (averaged over the six thermocouples) remains globally constant over time, despite some fluctuations. Consequently, we consider $\theta_{\text{gas}}$ to be proportional to $\theta$ by introducing $(1-\alpha)$, the proportionality factor, defined as follows: $\theta_{\text{gas}}=(1-\alpha)\theta$.
Hence, equation \eqref{bilanalu} leads to an exponential profile defined by the following equation:
\begin{equation}
    \Delta \theta(t)=\frac{\phi_{\text{in}}}{\alpha h}\left(1-e^{-t/\tau}\right),
    \label{solTalu}
\end{equation}
where $\Delta \theta(t)=\theta(t)-\theta_{\text{gas}}(t=0)$ and $\tau=\rho_{\text{Al}} c_{p_{\text{Al}}} e_{\text{Al}} /\alpha h$ represents the time constant of the system. 
\autoref{PT100plaque}b) illustrates the temporal evolution of the temperature at the center of the aluminum plate measured via the PT 100 probe, for different powers and bath temperatures. 
The green line on these profiles corresponds to a fit performed using the least squares method based on the solution \eqref{solTalu}. 
A very good agreement between this simple exponential model and the experimental measurements is observed. 
The fitting parameters $\Delta \theta_{\text{max}}$ (asymptotic value of $\Delta \theta(t)$ as $t\rightarrow \infty$) and $\tau$ then allow estimating the effective heat flux supplied to the system as $\phi_{\text{in}}=\rho_{\text{Al}} c_{p_{\text{Al}}} e_{\text{Al}}~\Delta \theta_{\text{max}}/\tau$ and consequently the heat efficiency coefficient $\chi$.


The second method, referred to as the \textit{steady method}, involves conducting an energy balance under steady state conditions.
Therefore, a thermal circuit representation of the different fluxes and temperatures can be drawn (as illustrated in \autoref{plate}).
The heat flux from the heating mats follows two paths. 
The main fraction conducts through the aluminum plate (thickness $e_{\text{Al}}$, conductivity $k_{\text{Al}}$), inducing convective flow in the system, characterized by the convective heat transfer coefficient $h$.
In a steady-state, $h$ can be determined using the thermal circuit (\autoref{plate}) as $h=\chi \phi_{\text{set}}/(\overline{\theta} -\overline{\theta}_{\text{gas}})$, where $\overline{\theta}$ and $\overline{\theta}_{\text{gas}}$ represent the steady-state temperatures of the aluminum plate and the gas within the system, respectively.
This path introduces a thermal resistance denoted as $R_{\text{Al}} = e_{\text{Al}}/{k_{\text{Al}}} + 1/{h}$.
The remaining heat travels through the insulating plate (thickness $e_{\text{IP}}$, conductivity $k_{\text{IP}}$) and the plexiglass support (thickness $e_{\text{plex}}$, conductivity $k_{\text{plex}}$) before contributing to small convective motion in the ambient air, characterized by the convective heat transfer coefficient $h_{\text{air}}$ \cite[determined via the  empirical free convection correlation of a hot horizontal plate heating from above, see e.g][]{Lloyd1974}.
This path is characterized by a thermal resistance noted as $R_{\text{ext}} = e_{\text{IP}}/k_{\text{IP}} +e_{\text{plex}}/{k_{\text{plex}}} + 1/h_{\text{air}}$.
Writing the global heat flux conservation, the heat efficiency coefficient can be expressed as  
\begin{equation}
    \chi = \frac{1}{1+\frac{R_{\text{Al}}}{R_{\text{ext}}}} - \frac{ \overline{\theta}_{\text{gas}} - \overline{\theta}_{\text{ext}}}{\phi_{\text{set}} \left(R_{\text{Al}}+R_{\text{ext}}\right) },
    \label{chisteady}
\end{equation}
with $\overline{\theta}_{\text{ext}}$ representing the room temperature.
The first term on the right-hand side corresponds to heat distribution due to thermal resistance ratio, while the second term represents a heat loss due to the temperature difference between the system and the ambient environment. 
It reflects the ratio between the flux due to the temperature difference across all layers (aluminum plate, insulated plate, etc.) and the flux set by the heating mats.
Notably, this term is relatively small compared to the first one, as $\phi_{\text{set}} \gg (\overline{\theta}_{\text{gas}} - \overline{\theta}_{\text{ext}})/(R_{\text{Al}}+R_{\text{ext}})$ and $R_{\text{ext}}$ is roughly 10 times larger than $R_{\text{Al}}$, making the first term the dominant factor.
We measured $\overline{\theta}$ using a PT100 on the aluminum plate and $\overline{\theta}_{\text{gas}}$ using the six thermocouples, during steady-state (after $3\tau$) and averaged these readings over time. 
These measurements provided the heat efficiency $\chi$.
In \autoref{heatlosses}, $\chi$ is plotted against the set power $P$ using both \textit{transient} and \textit{steady} methods. 
Both methods show good agreement, with an average thermal loss of \qty{15}{\%}. 
Error bars represent uncertainty propagation, calculated according to \eqref{solTalu} for the transient method and \eqref{chisteady} for the steady method, with contributions from the PT100 sensor uncertainty and the variances of the fit coefficients related to the aluminum plate temperature profiles.
Larger discrepancies at lower power levels are due to difficulties in fitting the exponential profile, where small temperature variations have a greater impact at lower power, leading to increased uncertainty. 
\begin{figure}
    \centering

 
\tikzset{
pattern size/.store in=\mcSize, 
pattern size = 5pt,
pattern thickness/.store in=\mcThickness, 
pattern thickness = 0.3pt,
pattern radius/.store in=\mcRadius, 
pattern radius = 1pt}
\makeatletter
\pgfutil@ifundefined{pgf@pattern@name@_ks5ddxj94}{
\pgfdeclarepatternformonly[\mcThickness,\mcSize]{_ks5ddxj94}
{\pgfqpoint{0pt}{0pt}}
{\pgfpoint{\mcSize+\mcThickness}{\mcSize+\mcThickness}}
{\pgfpoint{\mcSize}{\mcSize}}
{
\pgfsetcolor{\tikz@pattern@color}
\pgfsetlinewidth{\mcThickness}
\pgfpathmoveto{\pgfqpoint{0pt}{0pt}}
\pgfpathlineto{\pgfpoint{\mcSize+\mcThickness}{\mcSize+\mcThickness}}
\pgfusepath{stroke}
}}
\makeatother

 
\tikzset{
pattern size/.store in=\mcSize, 
pattern size = 5pt,
pattern thickness/.store in=\mcThickness, 
pattern thickness = 0.3pt,
pattern radius/.store in=\mcRadius, 
pattern radius = 1pt}
\makeatletter
\pgfutil@ifundefined{pgf@pattern@name@_pcfnjofl6}{
\makeatletter
\pgfdeclarepatternformonly[\mcRadius,\mcThickness,\mcSize]{_pcfnjofl6}
{\pgfpoint{-0.5*\mcSize}{-0.5*\mcSize}}
{\pgfpoint{0.5*\mcSize}{0.5*\mcSize}}
{\pgfpoint{\mcSize}{\mcSize}}
{
\pgfsetcolor{\tikz@pattern@color}
\pgfsetlinewidth{\mcThickness}
\pgfpathcircle\pgfpointorigin{\mcRadius}
\pgfusepath{stroke}
}}
\makeatother
\tikzset{every picture/.style={line width=0.75pt}} 

\begin{tikzpicture}[x=0.75pt,y=0.75pt,yscale=-1,xscale=1]

\draw    (327.67,3257.48) -- (327.67,3194.18) ;
\draw    (350.11,3257.48) -- (327.67,3257.48) ;
\draw [color={rgb, 255:red, 0; green, 0; blue, 0 }  ,draw opacity=1 ]   (294.06,3225.83) -- (285.54,3225.83) ;
\draw [shift={(285.54,3225.83)}, rotate = 180] [color={rgb, 255:red, 0; green, 0; blue, 0 }  ,draw opacity=1 ][fill={rgb, 255:red, 0; green, 0; blue, 0 }  ,fill opacity=1 ][line width=0.75]      (0, 0) circle [x radius= 3.35, y radius= 3.35]   ;
\draw    (495.95,3257.48) -- (487.89,3257.48) ;
\draw [shift={(495.95,3257.48)}, rotate = 180] [color={rgb, 255:red, 0; green, 0; blue, 0 }  ][fill={rgb, 255:red, 0; green, 0; blue, 0 }  ][line width=0.75]      (0, 0) circle [x radius= 3.35, y radius= 3.35]   ;
\draw  [color={rgb, 255:red, 255; green, 167; blue, 0 }  ,draw opacity=1 ] (406.6,3268.84) -- (362.51,3268.84) -- (362.51,3246.13) -- (406.6,3246.13) -- (406.6,3268.84) -- cycle (419,3257.48) -- (406.6,3257.48) (362.51,3257.48) -- (350.11,3257.48) ;
\draw [color={rgb, 255:red, 0; green, 0; blue, 0 }  ,draw opacity=1 ]   (294.06,3225.83) -- (325.67,3225.83) ;
\draw [shift={(327.67,3225.83)}, rotate = 180] [color={rgb, 255:red, 0; green, 0; blue, 0 }  ,draw opacity=1 ][line width=0.75]    (17.49,-5.26) .. controls (11.12,-2.23) and (5.29,-0.48) .. (0,0) .. controls (5.29,0.48) and (11.12,2.23) .. (17.49,5.26)   ;
\draw [color={rgb, 255:red, 0; green, 0; blue, 0 }  ,draw opacity=1 ]   (327.77,3257.48) -- (343.84,3257.5) ;
\draw [shift={(345.84,3257.51)}, rotate = 180.07] [color={rgb, 255:red, 0; green, 0; blue, 0 }  ,draw opacity=1 ][line width=0.75]    (13.12,-3.95) .. controls (8.34,-1.68) and (3.97,-0.36) .. (0,0) .. controls (3.97,0.36) and (8.34,1.68) .. (13.12,3.95)   ;
\draw [color={rgb, 255:red, 0; green, 0; blue, 0 }  ,draw opacity=1 ]   (327.67,3194.18) -- (348.11,3194.46) ;
\draw [shift={(350.11,3194.48)}, rotate = 180.79] [color={rgb, 255:red, 0; green, 0; blue, 0 }  ,draw opacity=1 ][line width=0.75]    (6.56,-1.97) .. controls (4.17,-0.84) and (1.99,-0.18) .. (0,0) .. controls (1.99,0.18) and (4.17,0.84) .. (6.56,1.97)   ;
\draw [color={rgb, 255:red, 245; green, 166; blue, 35 }  ,draw opacity=1 ]   (363.22,3153.64) -- (376.6,3153.64) ;
\draw [color={rgb, 255:red, 74; green, 144; blue, 226 }  ,draw opacity=1 ]   (458.05,3154.25) -- (471.02,3154.25) ;
\draw  [draw opacity=0][fill={rgb, 255:red, 74; green, 74; blue, 74 }  ,fill opacity=0.52 ] (91.64,3212.91) -- (233.16,3212.91) -- (233.16,3218.28) -- (91.64,3218.28) -- cycle ;
\draw  [draw opacity=0][fill={rgb, 255:red, 74; green, 74; blue, 74 }  ,fill opacity=0.52 ][line width=0.75]  (91.64,3217.86) -- (232.83,3217.86) -- (232.83,3212.91) -- (91.64,3212.91) -- cycle ;

\draw  [color={rgb, 255:red, 237; green, 237; blue, 237 }  ,draw opacity=1 ][fill={rgb, 255:red, 237; green, 237; blue, 237 }  ,fill opacity=1 ] (91.76,3219.47) -- (233.6,3219.47) -- (233.6,3228.74) -- (91.76,3228.74) -- cycle ;
\draw  [draw opacity=0][fill={rgb, 255:red, 237; green, 237; blue, 237 }  ,fill opacity=0.33 ] (91.87,3212.91) -- (233.39,3212.91) -- (233.39,3218.28) -- (91.87,3218.28) -- cycle ;
\draw  [color={rgb, 255:red, 74; green, 74; blue, 74 }  ,draw opacity=1 ][pattern=_ks5ddxj94,pattern size=6pt,pattern thickness=0.75pt,pattern radius=0pt, pattern color={rgb, 255:red, 74; green, 74; blue, 74}][line width=0.75]  (91.87,3217.86) -- (233.06,3217.86) -- (233.06,3212.91) -- (91.87,3212.91) -- cycle ;

\draw  [color={rgb, 255:red, 237; green, 237; blue, 237 }  ,draw opacity=1 ][pattern=_pcfnjofl6,pattern size=4.125pt,pattern thickness=0.75pt,pattern radius=0.75pt, pattern color={rgb, 255:red, 196; green, 196; blue, 196}] (91.76,3219.47) -- (233.75,3219.47) -- (233.75,3228.75) -- (91.76,3228.75) -- cycle ;
\draw [color={rgb, 255:red, 167; green, 23; blue, 23 }  ,draw opacity=1 ][line width=1.5]    (91.76,3218.35) -- (233.3,3218.35) ;
\draw [color={rgb, 255:red, 0; green, 0; blue, 0 }  ,draw opacity=1 ][line width=0.75]    (91.38,3229.64) -- (234.09,3229.64) ;
\draw [color={rgb, 255:red, 208; green, 2; blue, 27 }  ,draw opacity=1 ]   (163.5,3219.2) .. controls (161.84,3217.52) and (161.85,3215.85) .. (163.53,3214.2) .. controls (165.2,3212.54) and (165.21,3210.87) .. (163.56,3209.2) .. controls (161.91,3207.52) and (161.92,3205.85) .. (163.6,3204.2) .. controls (165.27,3202.54) and (165.28,3200.87) .. (163.63,3199.2) .. controls (161.98,3197.52) and (161.99,3195.85) .. (163.67,3194.2) .. controls (165.34,3192.54) and (165.35,3190.87) .. (163.7,3189.2) -- (163.71,3188.07) -- (163.76,3180.07) ;
\draw [shift={(163.78,3178.07)}, rotate = 90.39] [color={rgb, 255:red, 208; green, 2; blue, 27 }  ,draw opacity=1 ][line width=0.75]    (13.12,-3.95) .. controls (8.34,-1.68) and (3.97,-0.36) .. (0,0) .. controls (3.97,0.36) and (8.34,1.68) .. (13.12,3.95)   ;
\draw [color={rgb, 255:red, 208; green, 2; blue, 27 }  ,draw opacity=0.33 ]   (162.68,3224.1) .. controls (164.36,3225.76) and (164.37,3227.43) .. (162.71,3229.1) .. controls (161.05,3230.77) and (161.06,3232.44) .. (162.73,3234.1) .. controls (164.4,3235.76) and (164.41,3237.43) .. (162.76,3239.1) .. controls (161.1,3240.77) and (161.11,3242.44) .. (162.78,3244.1) -- (162.79,3245.85) -- (162.83,3253.85) ;
\draw [shift={(162.85,3255.85)}, rotate = 269.7] [color={rgb, 255:red, 208; green, 2; blue, 27 }  ,draw opacity=0.33 ][line width=0.75]    (6.56,-1.97) .. controls (4.17,-0.84) and (1.99,-0.18) .. (0,0) .. controls (1.99,0.18) and (4.17,0.84) .. (6.56,1.97)   ;
\draw [color={rgb, 255:red, 0; green, 0; blue, 0 }  ,draw opacity=1 ]   (163.5,3219.2) ;
\draw [shift={(163.5,3219.2)}, rotate = 0] [color={rgb, 255:red, 0; green, 0; blue, 0 }  ,draw opacity=1 ][fill={rgb, 255:red, 0; green, 0; blue, 0 }  ,fill opacity=1 ][line width=0.75]      (0, 0) circle [x radius= 3.35, y radius= 3.35]   ;
\draw  [color={rgb, 255:red, 255; green, 163; blue, 0 }  ,draw opacity=1 ] (406.6,3205.84) -- (362.51,3205.84) -- (362.51,3183.13) -- (406.6,3183.13) -- (406.6,3205.84) -- cycle (419,3194.48) -- (406.6,3194.48) (362.51,3194.48) -- (350.11,3194.48) ;
\draw  [color={rgb, 255:red, 74; green, 144; blue, 226 }  ,draw opacity=1 ] (475.49,3268.84) -- (431.4,3268.84) -- (431.4,3246.13) -- (475.49,3246.13) -- (475.49,3268.84) -- cycle (487.89,3257.48) -- (475.49,3257.48) (431.4,3257.48) -- (419,3257.48) ;
\draw  [color={rgb, 255:red, 255; green, 163; blue, 0 }  ,draw opacity=1 ] (475.49,3205.84) -- (431.4,3205.84) -- (431.4,3183.13) -- (475.49,3183.13) -- (475.49,3205.84) -- cycle (487.89,3194.48) -- (475.49,3194.48) (431.4,3194.48) -- (419,3194.48) ;
\draw  [color={rgb, 255:red, 74; green, 144; blue, 226 }  ,draw opacity=1 ] (544.38,3205.84) -- (500.29,3205.84) -- (500.29,3183.13) -- (544.38,3183.13) -- (544.38,3205.84) -- cycle (556.78,3194.48) -- (544.38,3194.48) (500.29,3194.48) -- (487.89,3194.48) ;
\draw    (564.83,3194.48) -- (556.78,3194.48) ;
\draw [shift={(564.83,3194.48)}, rotate = 180] [color={rgb, 255:red, 0; green, 0; blue, 0 }  ][fill={rgb, 255:red, 0; green, 0; blue, 0 }  ][line width=0.75]      (0, 0) circle [x radius= 3.35, y radius= 3.35]   ;

\draw (431.5,3187.78) node [anchor=north west][inner sep=0.75pt]  [font=\tiny,color={rgb, 255:red, 255; green, 163; blue, 0 }  ,opacity=1 ] [align=left] {\textcolor[rgb]{1,0.64,0}{{\scriptsize Plexiglas}}};
\draw (514.16,3187.95) node [anchor=north west][inner sep=0.75pt]  [font=\scriptsize] [align=left] {\textcolor[rgb]{0.29,0.56,0.89}{Air}};
\draw (375.67,3145.21) node [anchor=north west][inner sep=0.75pt]  [font=\scriptsize,color={rgb, 255:red, 255; green, 163; blue, 0 }  ,opacity=1 ] [align=left] {\textcolor[rgb]{1,0.64,0}{Diffusion}};
\draw (470.01,3144.99) node [anchor=north west][inner sep=0.75pt]  [font=\scriptsize] [align=left] {\textcolor[rgb]{0.29,0.56,0.89}{Convection}};
\draw (445.6,3252.24) node [anchor=north west][inner sep=0.75pt]  [font=\scriptsize] [align=left] {\textcolor[rgb]{0.29,0.56,0.89}{Gas}};
\draw (285.62,3168.75) node [anchor=north west][inner sep=0.75pt]  [font=\footnotesize]  {$( 1-\chi ) \phi _{\text{set}}$};
\draw (312.13,3259.92) node [anchor=north west][inner sep=0.75pt]  [font=\footnotesize]  {$\chi \phi _{\text{set}}$};
\draw (289.14,3198.81) node [anchor=north west][inner sep=0.75pt]    {$\phi _{\text{set}}$};
\draw (498.84,3242.8) node [anchor=north west][inner sep=0.75pt]    {$\overline{\theta}_{\text{gas}}$};
\draw (567.78,3179.89) node [anchor=north west][inner sep=0.75pt]    {$\overline{\theta}_{\text{ext}}$};
\draw (97.49,3186.95) node [anchor=north west][inner sep=0.75pt]    {$\overline{\theta}_{\text{gas}}$};
\draw (99.04,3233.94) node [anchor=north west][inner sep=0.75pt]    {$\overline{\theta}_{\text{ext}}$};
\draw (168.03,3238) node [anchor=north west][inner sep=0.75pt]  [font=\footnotesize]  {$( 1-\chi ) \ \phi _{\text{set}}$};
\draw (173.41,3170.7) node [anchor=north west][inner sep=0.75pt]  [font=\footnotesize]  {$\chi \ \phi _{\text{set}}$};
\draw (375.69,3192.14) node [anchor=north west][inner sep=0.75pt]  [font=\tiny,color={rgb, 255:red, 255; green, 163; blue, 0 }  ,opacity=1 ] [align=left] {\textcolor[rgb]{0.96,0.65,0.14}{{\scriptsize plate}}};
\draw (362,3183) node [anchor=north west][inner sep=0.75pt]  [font=\tiny,color={rgb, 255:red, 255; green, 163; blue, 0 }  ,opacity=1 ] [align=left] {\textcolor[rgb]{1,0.64,0}{{\scriptsize Insulated}}};
\draw (376.89,3253.74) node [anchor=north west][inner sep=0.75pt]  [font=\tiny] [align=left] {\textcolor[rgb]{1,0.64,0}{{\scriptsize plate}}};
\draw (379.06,3245.49) node [anchor=north west][inner sep=0.75pt]  [font=\tiny] [align=left] {\textcolor[rgb]{1,0.64,0}{{\scriptsize \ch{Al}}}};

\end{tikzpicture}
\caption{Left: sketch of the aluminum plate and the heat fluxes distribution due to the heating mats. Right: thermal circuit of the aluminum plate area, the orange and blue rectangles representing  diffusive and convective thermal resistances respectively. }
    \label{plate}
\end{figure}
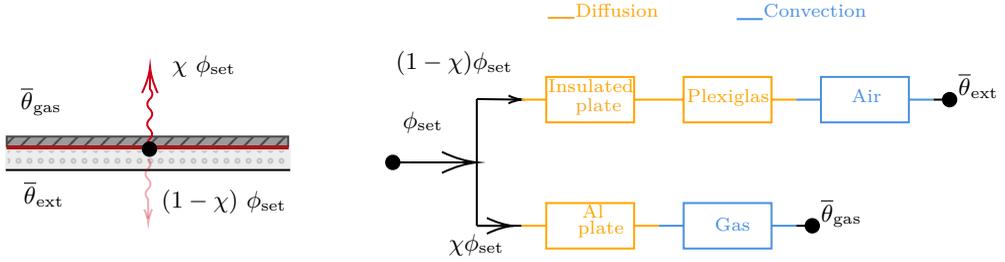
\begin{figure}
    \centering
    \includegraphics[scale=0.35]{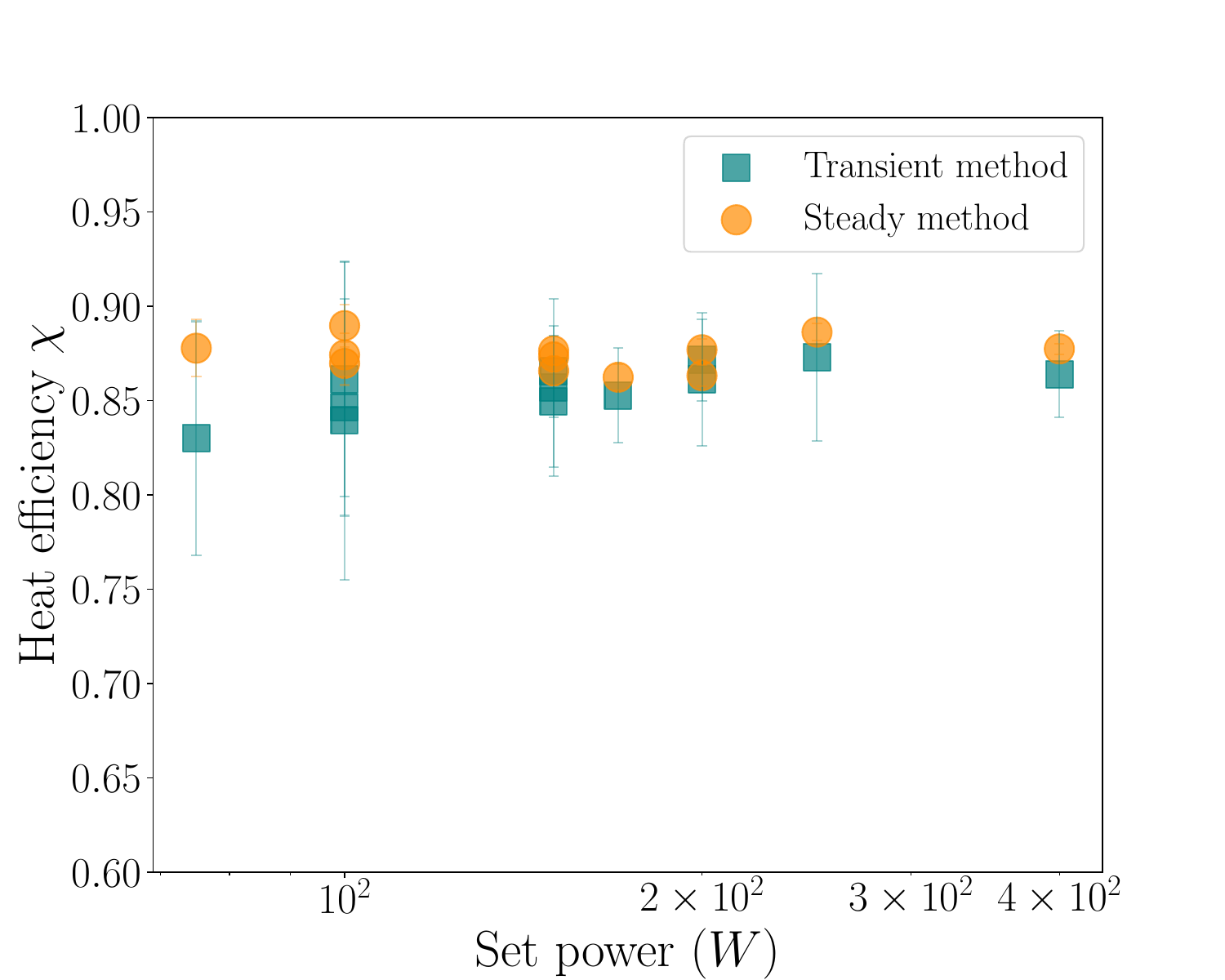}
   \caption{Heat efficiency coefficient $\chi = \phi_{\text{in}} / \phi_{\text{set}}$, as a function of the set power $P$ using the transient method ($\square$) and the steady method ($\circ$).}
    \label{heatlosses}
\end{figure}

\subsubsection{Flux ratio determination}
\label{estimationrf}
In our numerical model, a uniform and constant flux escaping from the upper surface is imposed. In the experimental setup, a rigid cover allows for a conductive heat flux, which is therefore non-uniform and fluctuating. This configuration resembles the real situation, where the resulting radiative flux is a dynamic consequence of the system's heat transfers. However, an average flux escaping from the upper surface can be estimated to evaluate a flux ratio $\RF$ and compare the results with numerical simulations. In this subsection, we outline two methods to estimate it. 
The room temperature is one of the parameter which can influence the top heat flux, similarly than the lid thickness and the top temperature in the system below the lid.
The room temperature is not controlled, but measured for each experiment and do not significatively change during the period of the data acquisitions ($\sim 1h$).

The first method, named the \textit{lid method}, involves performing a thermal balance on the lid.
To minimize disturbances to the flow, thin thermocouples (much thinner than PT-100) are carefully positioned above and just below the lid at the center to take the necessary measurements.
Through these measurements, the temperature difference between the bottom and top of the lid can be averaged and related to the flux escaping through conduction in the cover ($\RF \phi_{\text{in}}$) using the relation:
\begin{equation}
    \RF \phi_{\text{in}} = k_{\text{Lid}}~\frac{\Delta \theta_{\text{Lid}}}{e_{\text{Lid}}}
    \label{eqlid}
\end{equation}
where $ k_{\text{Lid}}$, $e_{\text{Lid}}$, $\Delta \theta_{\text{Lid}}$ represent the thermal conductivity, the thickness of the lid, and the temperature difference between the bottom and top of the lid, respectively.
Note that this flux ratio estimation is based on a local temperature measurement, so it does not account for system inhomogeneity.

\begin{figure}
     \centering


\tikzset {_fhgr0hkga/.code = {\pgfsetadditionalshadetransform{ \pgftransformshift{\pgfpoint{0 bp } { 0 bp }  }  \pgftransformrotate{-260 }  \pgftransformscale{2 }  }}}
\pgfdeclarehorizontalshading{_ux4sx33p3}{150bp}{rgb(0bp)=(0.65,0.81,0.87);
rgb(37.5bp)=(0.65,0.81,0.87);
rgb(62.5bp)=(0.75,0.88,1);
rgb(100bp)=(0.75,0.88,1)}

  
\tikzset {_vfyofplvo/.code = {\pgfsetadditionalshadetransform{ \pgftransformshift{\pgfpoint{89.1 bp } { -128.7 bp }  }  \pgftransformscale{1.32 }  }}}
\pgfdeclareradialshading{_qur9203g0}{\pgfpoint{-72bp}{104bp}}{rgb(0bp)=(1,1,1);
rgb(0bp)=(1,1,1);
rgb(25bp)=(0.93,0.93,0.93);
rgb(400bp)=(0.93,0.93,0.93)}

 
\tikzset{
pattern size/.store in=\mcSize, 
pattern size = 5pt,
pattern thickness/.store in=\mcThickness, 
pattern thickness = 0.3pt,
pattern radius/.store in=\mcRadius, 
pattern radius = 1pt}
\makeatletter
\pgfutil@ifundefined{pgf@pattern@name@_3xih5e0cs}{
\pgfdeclarepatternformonly[\mcThickness,\mcSize]{_3xih5e0cs}
{\pgfqpoint{0pt}{0pt}}
{\pgfpoint{\mcSize}{\mcSize}}
{\pgfpoint{\mcSize}{\mcSize}}
{
\pgfsetcolor{\tikz@pattern@color}
\pgfsetlinewidth{\mcThickness}
\pgfpathmoveto{\pgfqpoint{0pt}{\mcSize}}
\pgfpathlineto{\pgfpoint{\mcSize+\mcThickness}{-\mcThickness}}
\pgfpathmoveto{\pgfqpoint{0pt}{0pt}}
\pgfpathlineto{\pgfpoint{\mcSize+\mcThickness}{\mcSize+\mcThickness}}
\pgfusepath{stroke}
}}
\makeatother

  
\tikzset {_vbvgbj9q9/.code = {\pgfsetadditionalshadetransform{ \pgftransformshift{\pgfpoint{0 bp } { 0 bp }  }  \pgftransformrotate{-260 }  \pgftransformscale{2 }  }}}
\pgfdeclarehorizontalshading{_fetaxdfxp}{150bp}{rgb(0bp)=(0.65,0.81,0.87);
rgb(37.5bp)=(0.65,0.81,0.87);
rgb(62.5bp)=(0.75,0.88,1);
rgb(100bp)=(0.75,0.88,1)}

  
\tikzset {_8n06bv24m/.code = {\pgfsetadditionalshadetransform{ \pgftransformshift{\pgfpoint{89.1 bp } { -128.7 bp }  }  \pgftransformscale{1.32 }  }}}
\pgfdeclareradialshading{_ltoww1py3}{\pgfpoint{-72bp}{104bp}}{rgb(0bp)=(1,1,1);
rgb(0bp)=(1,1,1);
rgb(25bp)=(0.93,0.93,0.93);
rgb(400bp)=(0.93,0.93,0.93)}

  
\tikzset {_wlu49y24l/.code = {\pgfsetadditionalshadetransform{ \pgftransformshift{\pgfpoint{0 bp } { 0 bp }  }  \pgftransformscale{1 }  }}}
\pgfdeclareradialshading{_s869bcxtv}{\pgfpoint{0bp}{0bp}}{rgb(0bp)=(1,1,1);
rgb(0bp)=(1,1,1);
rgb(25bp)=(0.93,0.93,0.93);
rgb(400bp)=(0.93,0.93,0.93)}

  
\tikzset {_nrra6nq8k/.code = {\pgfsetadditionalshadetransform{ \pgftransformshift{\pgfpoint{89.1 bp } { -128.7 bp }  }  \pgftransformscale{1.32 }  }}}
\pgfdeclareradialshading{_5ra1diwkk}{\pgfpoint{-72bp}{104bp}}{rgb(0bp)=(1,1,1);
rgb(0bp)=(1,1,1);
rgb(25bp)=(0.93,0.93,0.93);
rgb(400bp)=(0.93,0.93,0.93)}
\tikzset{every picture/.style={line width=0.75pt}} 

  
\tikzset {_zwvk93qnc/.code = {\pgfsetadditionalshadetransform{ \pgftransformshift{\pgfpoint{37.5 bp } { 0 bp }  }  \pgftransformrotate{-260 }  \pgftransformscale{2 }  }}}
\pgfdeclarehorizontalshading{_pt8jer87f}{150bp}{rgb(0bp)=(0.14,0.33,0.54);
rgb(37.5bp)=(0.14,0.33,0.54);
rgb(62.5bp)=(0.65,0.81,0.87);
rgb(100bp)=(0.65,0.81,0.87)}

  
\tikzset {_1ttn8j64b/.code = {\pgfsetadditionalshadetransform{ \pgftransformshift{\pgfpoint{0 bp } { 0 bp }  }  \pgftransformrotate{-96 }  \pgftransformscale{6 }  }}}
\pgfdeclarehorizontalshading{_uq5r7oqfz}{150bp}{rgb(0bp)=(0.65,0.81,0.87);
rgb(37.5bp)=(0.65,0.81,0.87);
rgb(62.5bp)=(0.14,0.33,0.54);
rgb(100bp)=(0.14,0.33,0.54)}
\tikzset{every picture/.style={line width=0.75pt}} 

\begin{tikzpicture}[x=0.75pt,y=0.75pt,yscale=-1,xscale=1]


\path  [shading=_pt8jer87f,_zwvk93qnc] (282,3458.18) -- (370.33,3458.18) -- (370.33,3649.56) -- (282,3649.56) -- cycle ; 
 \draw  [color={rgb, 255:red, 74; green, 144; blue, 226 }  ,draw opacity=0.07 ] (282,3458.18) -- (370.33,3458.18) -- (370.33,3649.56) -- (282,3649.56) -- cycle ; 

\path  [shading=_qur9203g0,_vfyofplvo] (257.55,3428.67) -- (370.58,3428.67) -- (370.58,3459.02) -- (257.55,3459.02) -- cycle ; 
 \draw   (257.55,3428.67) -- (370.58,3428.67) -- (370.58,3459.02) -- (257.55,3459.02) -- cycle ; 

\draw  [color={rgb, 255:red, 74; green, 74; blue, 74 }  ,draw opacity=1 ][pattern=_3xih5e0cs,pattern size=3pt,pattern thickness=0.75pt,pattern radius=0pt, pattern color={rgb, 255:red, 0; green, 0; blue, 0}] (257.55,3670.9) -- (370.93,3670.9) -- (370.93,3681.59) -- (257.55,3681.59) -- cycle ;
\draw  [draw opacity=0][fill={rgb, 255:red, 139; green, 87; blue, 42 }  ,fill opacity=0.38 ] (257.55,3670.9) -- (371.26,3670.9) -- (371.26,3680.99) -- (257.55,3680.99) -- cycle ;

\path  [shading=_uq5r7oqfz,_1ttn8j64b] (233.29,3622.85) -- (282.09,3622.85) -- (282.09,3636.45) -- (233.29,3636.45) -- cycle ; 
 \draw  [color={rgb, 255:red, 74; green, 144; blue, 226 }  ,draw opacity=0.07 ] (233.29,3622.85) -- (282.09,3622.85) -- (282.09,3636.45) -- (233.29,3636.45) -- cycle ; 

\draw   (370.58,3459.02) -- (370.58,3649.56) -- (282,3649.56) -- (282,3636.83) -- (249.19,3636.83) -- (249.19,3636.7) -- (233.29,3636.7) -- (233.29,3622.85) -- (282,3622.85) -- (282,3459.02) -- (370.58,3459.02) -- cycle ;
\path  [shading=_ltoww1py3,_8n06bv24m] (257.55,3459.1) -- (282,3459.1) -- (282,3621.98) -- (257.55,3621.98) -- cycle ; 
 \draw   (257.55,3459.1) -- (282,3459.1) -- (282,3621.98) -- (257.55,3621.98) -- cycle ; 

\path  [shading=_s869bcxtv,_wlu49y24l] (257.55,3637.57) -- (282,3637.57) -- (282,3649.56) -- (257.55,3649.56) -- cycle ; 
 \draw   (257.55,3637.57) -- (282,3637.57) -- (282,3649.56) -- (257.55,3649.56) -- cycle ; 

\draw [color={rgb, 255:red, 245; green, 166; blue, 35 }  ,draw opacity=1 ][line width=1.5]    (414,3551.51) .. controls (412.33,3553.18) and (410.67,3553.18) .. (409,3551.51) .. controls (407.33,3549.84) and (405.67,3549.84) .. (404,3551.51) .. controls (402.33,3553.18) and (400.67,3553.18) .. (399,3551.51) .. controls (397.33,3549.84) and (395.67,3549.84) .. (394,3551.51) .. controls (392.33,3553.18) and (390.67,3553.18) .. (389,3551.51) .. controls (387.33,3549.84) and (385.67,3549.84) .. (384,3551.51) .. controls (382.33,3553.18) and (380.67,3553.18) .. (379,3551.51) .. controls (377.33,3549.84) and (375.67,3549.84) .. (374,3551.51) .. controls (372.33,3553.18) and (370.67,3553.18) .. (369,3551.51) .. controls (367.33,3549.84) and (365.67,3549.84) .. (364,3551.51) .. controls (362.33,3553.18) and (360.67,3553.18) .. (359,3551.51) -- (355.59,3551.51) -- (347.59,3551.51) ;
\draw [shift={(344.59,3551.51)}, rotate = 360] [color={rgb, 255:red, 245; green, 166; blue, 35 }  ,draw opacity=1 ][line width=1.5]    (19.89,-5.99) .. controls (12.65,-2.54) and (6.02,-0.55) .. (0,0) .. controls (6.02,0.55) and (12.65,2.54) .. (19.89,5.99)   ;
\draw [color={rgb, 255:red, 208; green, 2; blue, 27 }  ,draw opacity=0.4 ][line width=1.5]    (235.89,3551.93) .. controls (237.56,3550.26) and (239.22,3550.26) .. (240.89,3551.93) .. controls (242.56,3553.6) and (244.22,3553.6) .. (245.89,3551.93) .. controls (247.56,3550.26) and (249.22,3550.26) .. (250.89,3551.93) .. controls (252.56,3553.6) and (254.22,3553.6) .. (255.89,3551.93) .. controls (257.56,3550.26) and (259.22,3550.26) .. (260.89,3551.93) .. controls (262.56,3553.6) and (264.22,3553.6) .. (265.89,3551.93) .. controls (267.56,3550.26) and (269.22,3550.26) .. (270.89,3551.93) .. controls (272.56,3553.6) and (274.22,3553.6) .. (275.89,3551.93) .. controls (277.56,3550.26) and (279.22,3550.26) .. (280.89,3551.93) .. controls (282.56,3553.6) and (284.22,3553.6) .. (285.89,3551.93) -- (287.6,3551.93) -- (295.6,3551.93) ;
\draw [shift={(298.6,3551.93)}, rotate = 180] [color={rgb, 255:red, 208; green, 2; blue, 27 }  ,draw opacity=0.4 ][line width=1.5]    (12.79,-3.85) .. controls (8.13,-1.64) and (3.87,-0.35) .. (0,0) .. controls (3.87,0.35) and (8.13,1.64) .. (12.79,3.85)   ;
\draw [color={rgb, 255:red, 208; green, 2; blue, 27 }  ,draw opacity=0.25 ][line width=1.5]    (321.78,3416) .. controls (323.45,3417.67) and (323.45,3419.33) .. (321.78,3421) .. controls (320.11,3422.67) and (320.11,3424.33) .. (321.78,3426) .. controls (323.45,3427.67) and (323.45,3429.33) .. (321.78,3431) .. controls (320.11,3432.67) and (320.11,3434.33) .. (321.78,3436) .. controls (323.45,3437.67) and (323.45,3439.33) .. (321.78,3441) .. controls (320.11,3442.67) and (320.11,3444.33) .. (321.78,3446) .. controls (323.45,3447.67) and (323.45,3449.33) .. (321.78,3451) .. controls (320.11,3452.67) and (320.11,3454.33) .. (321.78,3456) -- (321.78,3459.25) -- (321.78,3467.25) ;
\draw [shift={(321.78,3470.25)}, rotate = 270] [color={rgb, 255:red, 208; green, 2; blue, 27 }  ,draw opacity=0.25 ][line width=1.5]    (8.53,-2.57) .. controls (5.42,-1.09) and (2.58,-0.23) .. (0,0) .. controls (2.58,0.23) and (5.42,1.09) .. (8.53,2.57)   ;
\draw [color={rgb, 255:red, 167; green, 23; blue, 23 }  ,draw opacity=1 ][line width=1.5]    (257.55,3681.59) -- (370.93,3681.59) ;
\draw    (285.69,3599.91) -- (366.71,3599.91) ;
\draw [shift={(368.71,3599.91)}, rotate = 180] [color={rgb, 255:red, 0; green, 0; blue, 0 }  ][line width=0.75]    (10.93,-3.29) .. controls (6.95,-1.4) and (3.31,-0.3) .. (0,0) .. controls (3.31,0.3) and (6.95,1.4) .. (10.93,3.29)   ;
\draw [shift={(283.69,3599.91)}, rotate = 0] [color={rgb, 255:red, 0; green, 0; blue, 0 }  ][line width=0.75]    (10.93,-3.29) .. controls (6.95,-1.4) and (3.31,-0.3) .. (0,0) .. controls (3.31,0.3) and (6.95,1.4) .. (10.93,3.29)   ;
\draw    (258.16,3483.98) -- (280.36,3483.98) ;
\draw [shift={(282.36,3483.98)}, rotate = 180] [color={rgb, 255:red, 0; green, 0; blue, 0 }  ][line width=0.75]    (6.56,-1.97) .. controls (4.17,-0.84) and (1.99,-0.18) .. (0,0) .. controls (1.99,0.18) and (4.17,0.84) .. (6.56,1.97)   ;
\draw [shift={(256.16,3483.98)}, rotate = 0] [color={rgb, 255:red, 0; green, 0; blue, 0 }  ][line width=0.75]    (6.56,-1.97) .. controls (4.17,-0.84) and (1.99,-0.18) .. (0,0) .. controls (1.99,0.18) and (4.17,0.84) .. (6.56,1.97)   ;
\draw    (350.58,3431) -- (350.58,3456.05) ;
\draw [shift={(350.58,3458.05)}, rotate = 270] [color={rgb, 255:red, 0; green, 0; blue, 0 }  ][line width=0.75]    (6.56,-1.97) .. controls (4.17,-0.84) and (1.99,-0.18) .. (0,0) .. controls (1.99,0.18) and (4.17,0.84) .. (6.56,1.97)   ;
\draw [shift={(350.58,3429)}, rotate = 90] [color={rgb, 255:red, 0; green, 0; blue, 0 }  ][line width=0.75]    (6.56,-1.97) .. controls (4.17,-0.84) and (1.99,-0.18) .. (0,0) .. controls (1.99,0.18) and (4.17,0.84) .. (6.56,1.97)   ;
\draw    (374.47,3673.19) -- (374.47,3681.1) ;
\draw [shift={(374.47,3683.1)}, rotate = 270] [color={rgb, 255:red, 0; green, 0; blue, 0 }  ][line width=0.75]    (4.37,-1.32) .. controls (2.78,-0.56) and (1.32,-0.12) .. (0,0) .. controls (1.32,0.12) and (2.78,0.56) .. (4.37,1.32)   ;
\draw [shift={(374.47,3671.19)}, rotate = 90] [color={rgb, 255:red, 0; green, 0; blue, 0 }  ][line width=0.75]    (4.37,-1.32) .. controls (2.78,-0.56) and (1.32,-0.12) .. (0,0) .. controls (1.32,0.12) and (2.78,0.56) .. (4.37,1.32)   ;
\draw    (254.57,3652.11) -- (254.57,3669.2) ;
\draw [shift={(254.57,3671.2)}, rotate = 270] [color={rgb, 255:red, 0; green, 0; blue, 0 }  ][line width=0.75]    (6.56,-1.97) .. controls (4.17,-0.84) and (1.99,-0.18) .. (0,0) .. controls (1.99,0.18) and (4.17,0.84) .. (6.56,1.97)   ;
\draw [shift={(254.57,3650.11)}, rotate = 90] [color={rgb, 255:red, 0; green, 0; blue, 0 }  ][line width=0.75]    (6.56,-1.97) .. controls (4.17,-0.84) and (1.99,-0.18) .. (0,0) .. controls (1.99,0.18) and (4.17,0.84) .. (6.56,1.97)   ;
\path  [shading=_5ra1diwkk,_nrra6nq8k] (257.55,3649.56) -- (370.58,3649.56) -- (370.58,3670.9) -- (257.55,3670.9) -- cycle ; 
 \draw   (257.55,3649.56) -- (370.58,3649.56) -- (370.58,3670.9) -- (257.55,3670.9) -- cycle ; 

\draw [color={rgb, 255:red, 208; green, 2; blue, 27 }  ,draw opacity=0.35 ][line width=1.5]    (323.62,3640) -- (323.62,3648) .. controls (325.29,3649.67) and (325.29,3651.33) .. (323.62,3653) .. controls (321.95,3654.67) and (321.95,3656.33) .. (323.62,3658) .. controls (325.29,3659.67) and (325.29,3661.33) .. (323.62,3663) .. controls (321.95,3664.67) and (321.95,3666.33) .. (323.62,3668) .. controls (325.29,3669.67) and (325.29,3671.33) .. (323.62,3673) .. controls (321.95,3674.67) and (321.95,3676.33) .. (323.62,3678) .. controls (325.29,3679.67) and (325.29,3681.33) .. (323.62,3683) .. controls (321.95,3684.67) and (321.95,3686.33) .. (323.62,3688) .. controls (325.29,3689.67) and (325.29,3691.33) .. (323.62,3693) .. controls (321.95,3694.67) and (321.95,3696.33) .. (323.62,3698) -- (323.62,3702.07) -- (323.62,3702.07) ;
\draw [shift={(323.62,3637)}, rotate = 90] [color={rgb, 255:red, 208; green, 2; blue, 27 }  ,draw opacity=0.35 ][line width=1.5]    (11.37,-3.42) .. controls (7.23,-1.45) and (3.44,-0.31) .. (0,0) .. controls (3.44,0.31) and (7.23,1.45) .. (11.37,3.42)   ;

\draw (301.56,3493.08) node [anchor=north west][inner sep=0.75pt]  [font=\large]  {$\theta _{\text{water}}$};
\draw (394.78,3494.34) node [anchor=north west][inner sep=0.75pt]  [font=\large]  {$\theta _{\text{gas}}$};
\draw (199.7,3494.19) node [anchor=north west][inner sep=0.75pt]  [font=\large]  {$\theta _{\text{ext}}$};
\draw (288.13,3686.14) node [anchor=north west][inner sep=0.75pt]  [font=\large]  {$\theta _{B}$};
\draw (254.28,3400.84) node [anchor=north west][inner sep=0.75pt]  [font=\small] [align=left] {\textcolor[rgb]{0.82,0.01,0.11}{External }\\\textcolor[rgb]{0.82,0.01,0.11}{heat flux }};
\draw (373.54,3596.05) node [anchor=north west][inner sep=0.75pt]  [color={rgb, 255:red, 139; green, 87; blue, 42 }  ,opacity=1 ] [align=left] {\textcolor[rgb]{0.96,0.65,0.14}{Lateral }\\\textcolor[rgb]{0.96,0.65,0.14}{heat flux}};
\draw (318.22,3579.03) node [anchor=north west][inner sep=0.75pt]    {$a$};
\draw (263.37,3485.95) node [anchor=north west][inner sep=0.75pt]    {$e_{S}$};
\draw (353,3440) node [anchor=north west][inner sep=0.75pt]    {$e_{T}$};
\draw (206.91,3532.41) node [anchor=north west][inner sep=0.75pt]  [font=\large,color={rgb, 255:red, 208; green, 2; blue, 27 }  ,opacity=1 ]  {$\phi ^{L}$};
\draw (324.06,3388.41) node [anchor=north west][inner sep=0.75pt]  [font=\large,color={rgb, 255:red, 208; green, 2; blue, 27 }  ,opacity=1 ]  {$\phi ^{T}$};
\draw (324.15,3683.74) node [anchor=north west][inner sep=0.75pt]  [font=\large,color={rgb, 255:red, 208; green, 2; blue, 27 }  ,opacity=1 ]  {$\phi ^{B}$};
\draw (375.82,3672.36) node [anchor=north west][inner sep=0.75pt]    {$e_{\text{cork}}$};
\draw (232.38,3655.05) node [anchor=north west][inner sep=0.75pt]    {$e_{B}$};
\draw (382.35,3558.06) node [anchor=north west][inner sep=0.75pt]  [font=\normalsize,color={rgb, 255:red, 139; green, 87; blue, 42 }  ,opacity=1 ]  {$\textcolor[rgb]{0.96,0.65,0.14}{(}\textcolor[rgb]{0.96,0.65,0.14}{1-R}\textcolor[rgb]{0.96,0.65,0.14}{_{F}}\textcolor[rgb]{0.96,0.65,0.14}{)}\textcolor[rgb]{0.96,0.65,0.14}{\ \phi _{\text{in}}}$};

\end{tikzpicture}
     \caption{Sketch representing the energy balance of the water within the torus.}
    \label{water}
\end{figure}

The second method, called the \textit{bath method}, relies on the thermal analysis of the cooling water on the lateral surface, providing a comprehensive measurement of the thermal fluxes involved in the system.
The thermal balance of the water within the torus can be formulated as 
\begin{equation}
    \underbrace{\rho_e c_{p_e} Q_v \Delta \theta_e}_\text{Change in water internal energy} = \underbrace{(1-\RF)\phi_{\text{in}}~S}_\text{Lateral flux} + \underbrace{\phi_{\text{ext}}~S_{\text{lat}}}_\text{External flux},
    \label{eqwater}
\end{equation}
leading to the average flux ratio 
\begin{equation}
\RF=1 -\left[\frac{\rho_e c_{p_e} Q_v \Delta \theta_e}{\phi_{\text{in}} S} - \frac{2\phi_{\text{ext}}}{\Gamma \phi_{\text{in}}}\right].
\end{equation}
Here, $\rho_e$, $c_{p_e}$, $Q_v$, $S_{\text{lat}}$, $S$ and $\Delta \theta_e$ represent the density, specific heat capacity, volumetric flow rate of water, the lateral ($S_{\text{lat}}=2\pi R H$) and the heating surface area ($S=\pi R^2$), and the temperature difference between water inlet and outlet in the bath, respectively.

Two PT 100 sensors enable the measurement of $\Delta \theta_e$. Additionally, a volumetric water meter measures the average flow rate of the system over the entire experiment duration ($\sim 12$ hours).
Equation \eqref{eqwater} indicates that the water undergoes a change in internal energy as it passes through the torus, due to the lateral heat flux of the system and the perturbative fluxes induced by the environment.
In equation \eqref{eqwater}, the lateral heat flux is expressed as the difference between the flux imposed on the bottom surface ($\phi_{\text{in}} S$) and that escaping from the cover ($R_F \phi_{\text{in}} S$). The term $\phi_{\text{ext}}$ denotes the perturbative heat flux from the external environment, heating the bath water. 
As sketched in \autoref{water}, it decomposes into three contributions $\phi^L$, $\phi^T$, and $\phi^B$, representing the external heat fluxes from the lateral surface, the upper surface of the torus, and the lower surface of the torus due to the heating elements:
\begin{equation}
\phi_{\text{ext}}= \phi^L+ (\phi^T+\phi^{B})\frac{a}{H}\left(1+\frac{a}{2R}\right).
\end{equation}
Here, $a$ represents the gap between the inner and outer walls of the torus. 
The perturbative fluxes can be expressed in terms of the torus characteristics and system temperatures as follows
\begin{gather}
\phi^L=\left(\frac{e_L}{k_{\text{PMMA}}} +\frac{1}{h_{\text{ext}}}\right)^{-1}\left(\theta_{\text{water}} - \theta_{\text{ext}}\right)\\
\phi^T=\left(\frac{e_T}{k_{\text{PMMA}}} +\frac{1}{h_{\text{ext}}}\right)^{-1}\left(\theta_{\text{water}} - \theta_{\text{ext}}\right)\\
\phi^{B}=\left(\frac{e_{\text{cork}}}{k_{\text{cork}}} +\frac{e_{B}}{k_{\text{PMMA}}}\right)^{-1}\left(\theta_{\text{water}} - \theta_{B} \right).
\end{gather}
where $k_{\text{PMMA}}$, $e_B$, $e_L$, $e_T$, $\theta_{\text{water}}$, $\theta_{\text{ext}}$, $\theta_{B}$ represent respectively the thermal conductivity of PMMA, the thickness of the bottom, side, and top of the torus, the temperature of the water, external environment, and of the contact between the heating element and the cork.
The convective heat transfer coefficient $h_{\text{ext}}$ characterizes the vertical convection generated by the external side wall and external air.
While strictly speaking, separate $h$ values apply to the vertical surface and the flat top, their estimated values are of the same order of magnitude (see e.g {\cite{FUJII1972}}).
Therefore, we use the same $h$ value for both, based on the classical vertical convection correlation \citep{churchill}.

During the measurement campaign, consistent efforts were made to maintain a temperature difference of approximately 5°C between the bathwater and the external environment. 
This approach was intended to keep the same external perturbative flux across all experiments.
An estimation of the perturbative term $\phi_{\text{ext}}$ was performed and represents an average of approximately \qty{12}{\%} of the imposed flux across all experiments.

\begin{figure}
    \centering
    \includegraphics[scale=0.35]{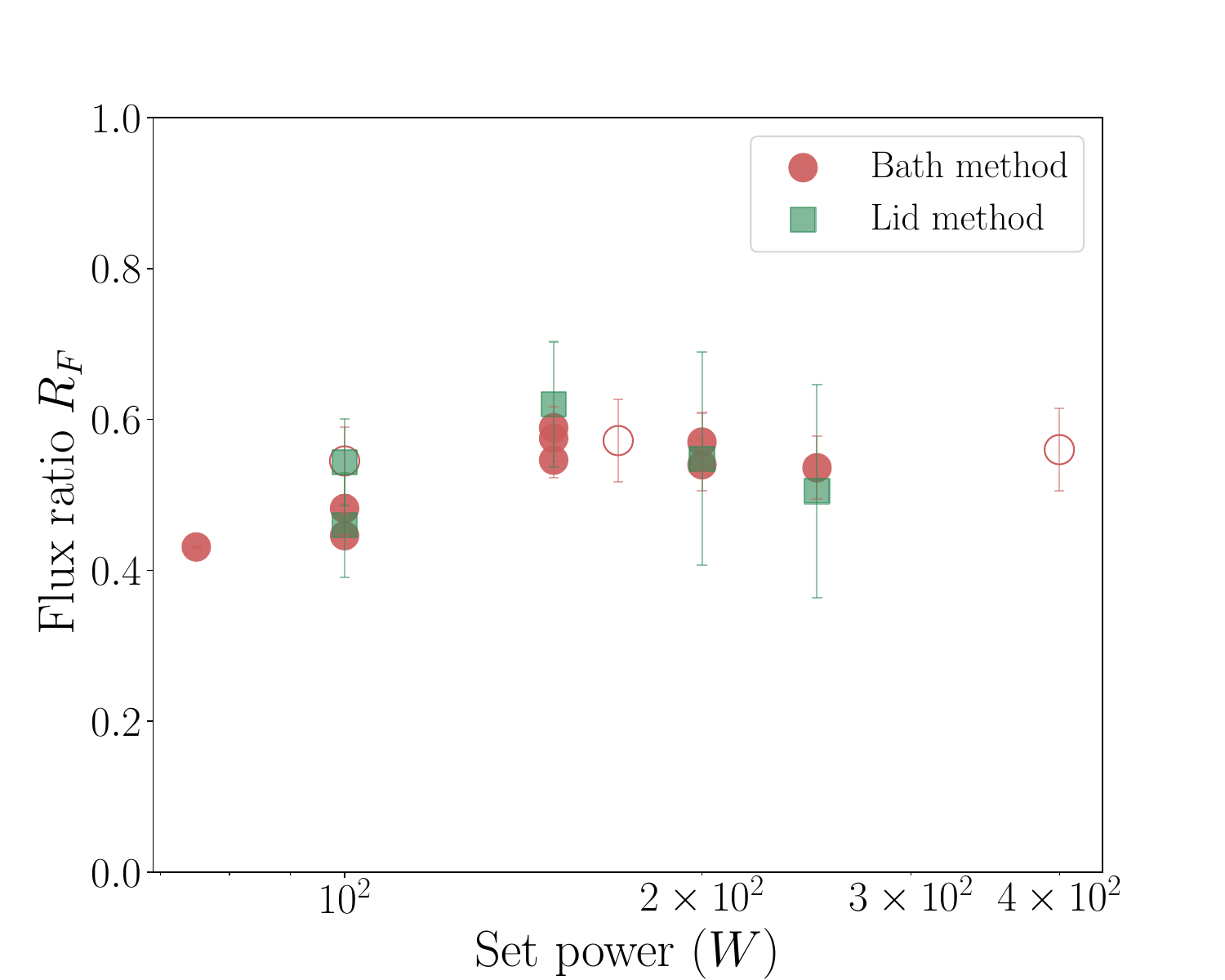}
   \caption{Estimated flux ratio $\RF$ as a function of the set power, considering Lid ($\square$) and Bath ($\circ$) methods. The empty symbols indicate \ch{SF6}  gas is used.}
    \label{rf}
\end{figure}

In \autoref{rf}, the estimation of the flux ratio using either the global method (\textit{bath method}) or the local method (\textit{lid method}) is plotted against the power supplied to the heating elements.
Error bars represent uncertainty propagation, calculated according to \eqref{eqlid} for the lid method and \eqref{eqwater} for the bath method, with contributions from the PT100, thermocouples uncertainties, as well as $\phi_{\mathrm{in}}$ uncertainty related to the variances of the fit coefficients from \eqref{solTalu}.
It is observed that the average flux ratio depends on the set power, ranging from approximately 0.4 to around 0.6. The global measurement using the bath provides a good experimental repeatability, with a measurement uncertainty of 0.02 for the measurement at \qty{150}{W}. Furthermore, the local measurements using the lid are consistent and in good agreement with the global ones. 
\bibliographystyle{jfm}
\bibliography{jfm-instructions}
\end{document}